# Structure and Stability of the Iodide Elpasolite, $Cs_2AgBiI_6$

Kyle T. Kluherz,[1] Sebastian T. Mergelsberg,[2] James J. De Yoreo,[2] and Daniel R. Gamelin[1,*]

[1]*Department of Chemistry, University of Washington, Seattle, WA 98195, U.S.A.*

[2]*Physical Sciences Division, Pacific Northwest National Laboratory,*

*Richland, WA 99352, U.S.A.*

Email: gamelin@chem.washington.edu

**Abstract.** Iodide elpasolites (or double perovskites, $A_2B'B''I_6$, $B' = M^+$, $B'' = M^{3+}$) are predicted to be promising alternatives to lead-based perovskite semiconductors for photovoltaic and optoelectronic applications, but no iodide elpasolite has ever been definitively prepared or structurally characterized. Iodide elpasolites are widely predicted to be unstable due to favorable decomposition to the competing $A_3B_2I_9$ ($B = M^{3+}$) phase. Here, we report the results of synchrotron XRD and X-ray total scattering measurements on putative $Cs_2AgBiI_6$ nanocrystals made *via* anion exchange from parent $Cs_2AgBiBr_6$ nanocrystals. Rietveld refinement of XRD and PDF data shows that these nanocrystals indeed exhibit a tetragonal (I4-m) elpasolite structure, making them the first example of a structurally characterized iodide elpasolite. A series of experiments probing structural relaxation and the effects of surface ligation or grain size all point to the critical role of surface free energy in stabilizing the iodide elpasolite phase in these nanocrystals.

**Introduction**

Lead-halide perovskites ($APbX_3$) have been extensively studied due to their excellent optoelectronic properties and many potential applications, including photovoltaics, lighting, and X-ray detection.[1-3] Recently, attention has turned to the possibility that elpasolites (or double perovskites, $A_2B'B''X_6$, where $B' = M^+$, $B'' = M^{3+}$) may serve as promising lead-free alternatives to the lead-halide perovskites.[4] Although elpasolites generally do not exhibit the excellent emissive properties of certain lead-halide perovskites, some do show long carrier lifetimes and broad absorption, making them potentially suitable for applications in photovoltaics.[4-7] Some elpasolites also exhibit superior thermal and water stability compared to lead-halide perovskites, making them potentially attractive for devices that experience high temperatures under standard operation.[8-11] Numerous theoretical studies of elpasolites have predicted that iodide elpasolites, in particular, should exhibit the most suitable band gaps for photovoltaic applications.[5,12-18]

Despite this broad interest and motivation, iodide elpasolites remarkably remain almost entirely unexplored experimentally. In fact, a recent review article[12] has highlighted that to date there have been *no* "structurally characterized" iodide elpasolites reported at all, asserting that their successful synthesis has so far been elusive. Computational models generally predict that iodide elpasolites are thermodynamically unstable relative to competing phases, explaining the paucity of experimental results. Although their enthalpies of formation may be favorable (*e.g.*, $\Delta H_f$ = -0.86 eV/atom for 2CsI + BiI$_3$ + AgI → Cs$_2$AgBiI$_6$),[19] they also appear to suffer from favorable decomposition to the A$_3$B$_2$X$_9$ (B = M$^{3+}$) or A$_3$BX$_6$ phases[12,19] (*e.g.*, $\Delta H_f$ = -0.41 eV/atom for 2Cs$_2$AgBiI$_6$ → Cs$_3$Bi$_2$I$_9$ + 2AgI + CsI).[5] The XRD patterns of these primary competing phases can appear very similar to those of the elpasolites, complicating experimental analysis.[9,20,21] We note that the thermodynamic properties of elpasolite *nanocrystals* likely differ significantly from those estimated for bulk, but to our knowledge, no theoretical investigations have yet examined the stabilities of nanoscale elpasolites.

Nevertheless, syntheses of a few iodide elpasolites have been claimed, suggesting that it may indeed be possible to stabilize this family of materials under specific circumstances. Cs$_2$NaLaI$_6$ was reported by Glodo *et al*[22] in 2006 and by Gundiah *et al.*[23] in 2014, but the lack of structural characterization leaves unclear whether the material was actually an elpasolite. Zhang *et al.* reported powder XRD (pXRD) data for Cs$_2$NaBiI$_6$ in 2018,[24] along with Zheng *et al.* in 2022,[25] but in both cases the clear presence of a Cs$_3$Bi$_2$I$_9$ contaminant and the strongly overlapping peaks of the two phases made definitive structural identification uncertain. Rb$_2$AgBiI$_6$ was reported by Bhorde *et al* in 2021,[26] but with poor agreement between the experimental pXRD and the diffraction pattern calculated from a DFT model. Shadabroo *et al.*[27] reported formation of MA$_2$AgBiI$_6$ in 2021, and Cs$_2$AgBiI$_6$ nanocrystals were reported by our group[9] and by Yang *et al.*[28] in 2018, but these materials were only structurally characterized with pXRD data that was insufficient to unambiguously identify them as iodide elpasolites. Notably, the absorption spectrum of Cs$_2$AgBiI$_6$ reported in ref. 28 is different from that reported in ref. 9 but is indistinguishable from that of one polymorph of Cs$_3$Bi$_2$I$_9$ (ref. 20, *vide infra*). As a cautionary note, a thorough structural study of another potential iodide elpasolite, Cu$_2$AgBiI$_6$, using 100 K single-crystal XRD measurements found this material to exhibit a *non*-elpasolite trigonal R3-m structure with layers of 2D edge-sharing octahedra alternating with vacant octahedral sites.[21,29] The authors proposed that this alternating layer structure provides stability over the elpasolite structure in this case. Overall, there is thus little to no unambiguous evidence for the existence of any *bone fide* iodide elpasolites, making this family a remarkable void in the composition space of this important class of materials.

Here, we follow up on our previous claim[9] of $Cs_2AgBiI_6$ elpasolite nanocrystals by reporting rigorous structural characterization of these nanocrystals *via* Rietveld refinement of synchrotron high-energy XRD (heXRD) data, analyzed in tandem with X-ray total scattering data. The data and analysis presented here yield the first unambiguous structure model of an iodide elpasolite, showing these nanocrystals to adopt the tetragonal I4-m structure. We further demonstrate the critical role of nanostructuring, and especially of surface ligation, in stabilizing this phase and preventing decomposition to the competing $Cs_3Bi_2I_9$ product. These results firmly establish the existence of iodide elpasolites, a family of materials predicted to excel in optoelectronics applications but never previously available for experimental investigation. More broadly, these results highlight the power of chemistries unique to the nanoscale for accessing unprecedented compositions of matter.

**Experimental Section**

**Materials.** $Bi(OAc)_3$, $Ag(OAc)$, $Cs(OAc)$, trimethylsilyl iodide (TMSI), trimethylsilyl bromide (TMSBr), octadecene (90%), oleic acid (OA) (90%), oleylamine (OLA) (70%), 3-(N,N-dimethyl-octadecylammonio)-propanesulfonate (sulfobetaine, >99%), benzyl alcohol (anhydrous, 99.8%), didodecyl-dimethylammonium bromide (DDDMABr), and trioctylphosphine (TOP, anhydrous, 90%) were purchased from Sigma Aldrich and used without further purification. Hydrated reagents (OA, OLA, sulfobetaine, DDDMABr) were dried under vacuum for 4 h then transferred into a glovebox before use in small-molecule addition experiments.

**Synthesis and anion exchange.** $Cs_2AgBiBr_6$ nanocrystals (NCs) were synthesized using the procedure we reported previously (see ref. 9 for details). $Cs_2AgBiI_6$ NCs were made from these $Cs_2AgBiBr_6$ NCs *via* anion exchange using TMSI.[9,30] In a standard complete anion-exchange reaction, 1 mL of neat TMSI was added to 5 mL of a ~10 μM $Cs_2AgBiBr_6$ NC solution in hexanes and the reaction vial sealed for 1 day. The solution changed color from yellow to dark brown-red within ~10 s, but was allowed to sit to ensure complete anion exchange. Absorption, XRD, and EDX data were collected to monitor reaction progress. Thin-film samples of $Cs_2AgBiBr_6$ were prepared by thermal evaporation as described previously.[31] Films were initially characterized in ambient atmosphere. For anion-exchange reactions, films were exposed to aliquots of TMSI vapor in a sealed glass vessel under inert atmosphere. For sequential TMSI exposures, the vessel's atmosphere was flushed with nitrogen between exposures. For thin-film heating experiments, samples (on Si and glass substrates) were heated directly on a temperature-controlled hot plate in a nitrogen glovebox. For NC heating experiments, colloidal samples were heated in a sealed glass vessel to minimize hexane evaporation.

**Small-molecule additions.** Small-molecule additions were performed in the following way: 5, 10, 20, or 100 μmol of the selected compound (OA, OLA, benzyl alcohol, sulfobetaine, DDDMABr, and TOP) was added to 0.5 mL of 10 μM $Cs_2AgBiX_6$ (X = Br, I) NC solutions in hexanes, briefly stirred, then allowed to react overnight (~20 h). With the exception of sulfobetaine, which was added as a solid powder, 0.1 M solutions of each compound in hexanes were used for these reactions. We primarily report results of the 20 μmol additions. After additions, $Cs_2AgBiBr_6$ samples were characterized using UV-Vis absorption, XRD, and TEM, then exposed to 100 μL TMSI for anion exchange.

**Sample characterization.** Absorption spectra were collected using an Agilent Cary 60. Benchtop powder XRD data were measured using a Bruker D8 Discover with a high-efficiency IμS microfocus X-ray source for Cu Kα radiation operating at 50,000 mW (50 kV, 1 mA). NC samples were prepared by drop-casting NC stock solutions on silicon substrates. Samples were measured under Kapton film to prevent air exposure during measurement. TEM images were acquired using an FEI Tecnai G2 F20 supertwin microscope operating at 200 kV. A C2 aperture of 70 μm was used to minimize beam damage. TEM EDX measurements were acquired using an EDAX-Elite-T detector. TEM measurements with *in situ* heating were performed using a FEI Titan 80-300 Environmental Transmission Electron Microscope (ETEM) with image correction. A Gatan double-tilt, furnace-based heating holder (Model 646) was used for sample heating. Samples were first imaged, then heated to 100 or 120°C for 10-30 min, then cooled to 30°C before further imaging to avoid beam damage. NC stock solutions (10 μM) were diluted by about one-fifth, and 5 μL of the solution was deposited onto ultrathin carbon Type A 400 Cu grids from TED Pella, Inc. SEM images were taken using an Apreos-S, and EDX mapping was performed at 10 kV, 800 pA, with a 100 ms dwell time.

**PDF measurements.** X-ray total scattering data were collected using beamline 11-ID-B at the Advanced Photon Source at Argonne National Laboratory.[32] Solutions of NC samples were measured at room temperature in quartz or Kapton capillaries using monochromatic X-rays with energy ~86.7 keV ($\lambda$ = 0.1432 Å). Using a Perkin Elmer 1621 a-Si area detector (200 μm$^2$ pixel size), the sample-to-detector distance and detector non-orthogonality were calibrated with a $CeO_2$ standard (NIST 674a) diluted with glassy carbon. X-ray total scattering data were integrated using the GSAS-II software[33] and a radial bin size of 1396, 1200, or 1000 for sample-to-detector distances of 250, 1000, and 1500 mm, respectively. Background subtraction and PDF processing were performed using the PDFgetX3 software.[34] For PDF processing, a $Q_{max}$ of 28 Å$^{-1}$ and $r_{max}$ of 30 Å were used. The PDFgui package[35] was used to determine the instrument parameters of $q_{damp}$ = 0.025 and $q_{broad}$ = 0.047, and to run r-space refinements to the PDF data. Rietveld refinements were calculated using the GSAS-II software,[33] with an analytic Hessian algorithm run for 3 cycles for each refinement. Unit cell, background (chebyschev-1 function, 9 to 15 terms), atom positions, and thermal parameters were refined. Fourier transforms of the backgrounds show no significant contribution of the elpasolite or secondary crystalline phases (Fig. S4).

**Results and Analysis**

**A. Structural characterization.** Figure 1 shows XRD and TEM data collected for $Cs_2AgBiI_6$ NC samples prepared by the method we reported previously.[9] As noted in ref. [9], $Cs_2AgBiI_6$ NCs will not form by direct hot-injection synthesis under our reaction conditions, and they are instead prepared *via* anion exchange from parent $Cs_2AgBiBr_6$ NCs that can be synthesized directly. The XRD data in Fig. 1a are consistent with the predicted XRD pattern for the $Cs_2AgBiI_6$ elpasolite phase (Fig. S2),[9] and notably show no identifiable peaks from the competing $Cs_3Bi_2I_9$ phase. The TEM data in Fig. 1b show cubic or near-cubic particles with an average edge length of ~14 nm (Fig. 1c) measured across ~200 particles. Overall, these data are essentially indistinguishable from the results we reported previously, demonstrating reproducibility of the synthesis methods from Creutz *et al*,[9] and indicating that the findings of this study are applicable to the materials reported in that previous study.

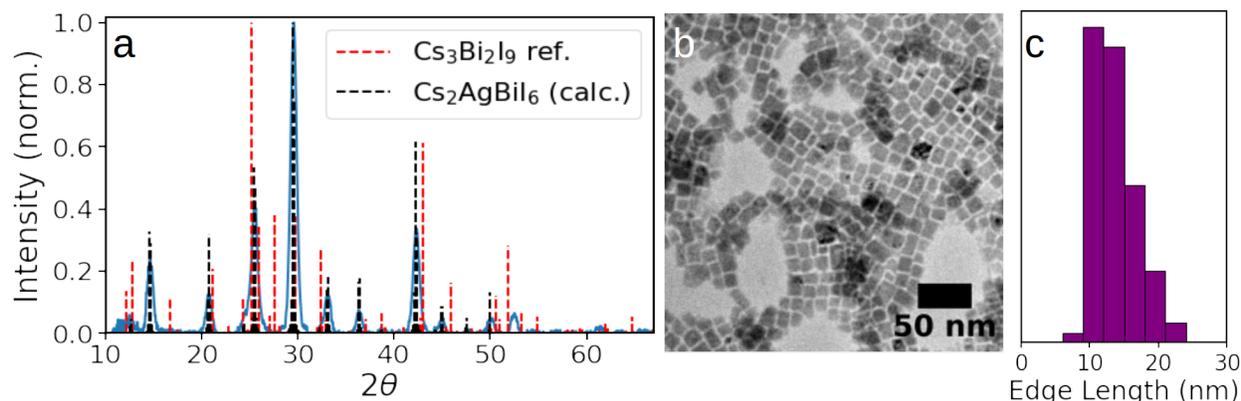

**Figure 1. (a)** Powder XRD data collected for $Cs_2AgBiI_6$ nanocrystals prepared by anion exchange from $Cs_2AgBiBr_6$ nanocrystals. A calculated pattern based on a hypothesized cubic (Fm-3m) $Cs_2AgBiI_6$ unit cell was used for initial identification. $Cs_3Bi_2I_9$ was selected for comparison as the most similar known competing phase. **(b)** TEM image of $Cs_2AgBiI_6$ nanocrystals. **(c)** Histogram of $Cs_2AgBiI_6$ nanocrystal sizes measured from TEM images of ~200 nanocrystals.

Although compelling, the XRD data in Fig. 1a are insufficient to unambiguously determine the structure of these $Cs_2AgBiI_6$ NCs. In particular, the data do not allow distinction among the three variants of elpasolites, the trigonal (R-3m), cubic (Fm-3m), and tetragonal (I4-m) structures. More powerful structural data were therefore obtained. Figure 2 shows high-energy X-ray scattering data collected for the same $Cs_2AgBiI_6$ NCs using a synchrotron source at the APS. Using these data, we calculated Rietveld refinement structure fits to the cubic, tetragonal, and trigonal elpasolite structures using GSAS-II.[33] The trigonal structural model yielded a poor fit with an overall weighted R factor ($R_w$) of 13.7 (see Fig. S6) and was not considered further. The cubic Fm-3m structural model gave a better fit, with an $R_w$ of 7.8 and low residual values (see Fig. S3), but inspection of the computed structure revealed unusually large thermal disorder within the $Ag^+$ sublattice (Fig. S3c). Fixing these thermal parameters to the literature values found for $Ag^+$ in $Cs_2AgBiBr_6$ (ICSD# 239875), which is known to be cubic (Fm-3m),[36] yielded a substantially poorer fit (Fig. S5), with $R_w$ = 12.9, and generated notable new residual peaks around 1.8 Å$^{-1}$ and 3.1 Å$^{-1}$. Close inspection found these large residuals to correlate with low-symmetry splittings of the (3,1,1) and (5,3,1) peaks, confirming a lower-symmetry structure. Rietveld refinement using a tetragonal I4-m structure (Fig. 2a, b) yielded a much improved fit ($R_w$ = 6.7) to the data. Notably, the lower $R_w$ relative to the Fm-3m fit is also accompanied by thermal parameters (Fig. 2c) very similar to those found in other elpasolite structures in the ICSD (Coll. Codes 239875, 11523, 18989, 21475, 32193). We therefore conclude that our $Cs_2AgBiI_6$ NCs are best described by a tetragonal I4-m elpasolite structure. This structure (Fig. 2b) is characterized by lattice parameters of $a = b = 8.535(4)$ Å and $c = 12.080(4)$ Å, a unit cell volume of 880 Å$^3$, and a distorted Bi-I octahedron with Bi-I bond distances of 2.91 (4 bonds) and 3.03 Å (2 bonds). The full structural details are cataloged in the accompanying cif file (see SI) and have been deposited in the Inorganic Crystal Structure Database hosted by the Cambridge Crystallographic Data Centre (CCDC). To our knowledge, these results represent the first complete structural characterization of any iodide elpasolite.

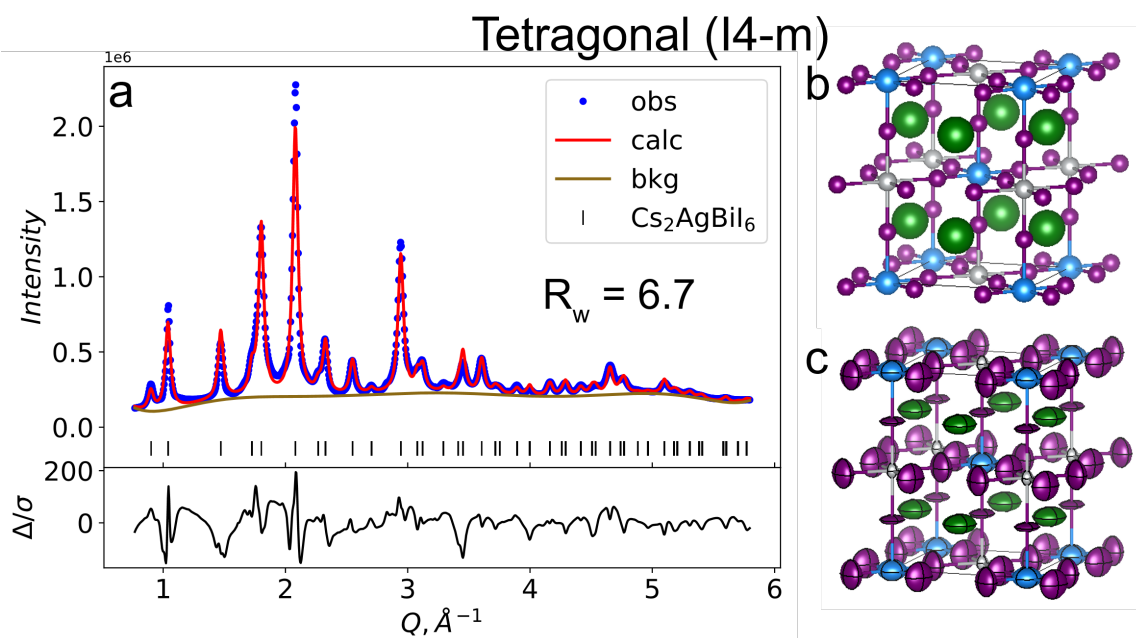

**Figure 2.** Rietveld refinement results using synchrotron X-ray scattering data for $Cs_2AgBiI_6$ nanocrystals suspended in hexanes solution. **(a)** Data (blue dots), calculated pattern (red curve), background (brown curve), predicted peak positions (vertical black lines), and residual of fit ($\Delta/\sigma$) obtained using tetragonal I4-m structure. **(b)** Tetragonal I4-m structure model. Silver: $Ag^+$; blue: $Bi^{3+}$; green: $Cs^+$; purple: $I^-$. **(c)** Structure model produced from (b), showing 95% atomic-displacement parameters refined using PDF data.

To supplement the above analysis of the $Cs_2AgBiI_6$ structure, we additionally investigated the structure's evolution during conversion of the NCs from cubic $Cs_2AgBiBr_6$ to tetragonal $Cs_2AgBiI_6$ by anion exchange. Figure 3 presents the heXRD and pair distribution functions (PDF) from X-ray total scattering data collected for a series of intermediate anion-exchange reactions. Dashed lines in Fig. 3b trace the progression of select peak positions across the sample series for ease of viewing. The first two peaks correspond to specific atom pairs and both show gradual shifts to larger r. The third is included to demonstrate that the same shift occurs in the medium-range order. Figure 3c,d summarizes these data by plotting formula unit volume ($V_f$) and PDF peak positions *vs* the fractional iodide content *x* in $Cs_2AgBi(Br_{1-x}I_x)_6$ measured by EDX. These data all increase linearly with *x*, reflecting lattice accommodation of the larger iodide anions. Both the heXRD and the PDF (Fig. 3a,b) data thus reveal continuous shifts in feature positions across the entire sample set, with no abrupt or discontinuous transitions between the two end points. PDF is especially useful in analysis of this conversion due to its high sensitivity to the presence of minority phases, which appear as new peaks or shoulders in G(r) regardless of crystallinity.[37-39] Notably, the data in Fig. 3b show no new peaks emerging during anion exchange, again supporting a gradual structural evolution with anion exchange. The data in Fig. 3 thus show no evidence of mixed-phase or exsolved-phase compositions at any of the intermediate stages of anion exchange, and we conclude that these materials transition continuously from the cubic $Cs_2AgBiBr_6$ structure to the tetragonal $Cs_2AgBiI_6$ structure *via* incremental lattice expansion during anion exchange. This conclusion is similar to that drawn for lead-halide perovskite NCs ($CsPbX_3$), which also convert from chloride to bromide to iodide compositions with no detectable intermediate or exsolved phases.[30,40-42] Importantly, the gradual evolution of these data in conjunction with the fact that $Cs_2AgBiBr_6$ is a well-established elpasolite structure bolsters our conclusion that $Cs_2AgBiI_6$ retains an elpasolite structure as well.

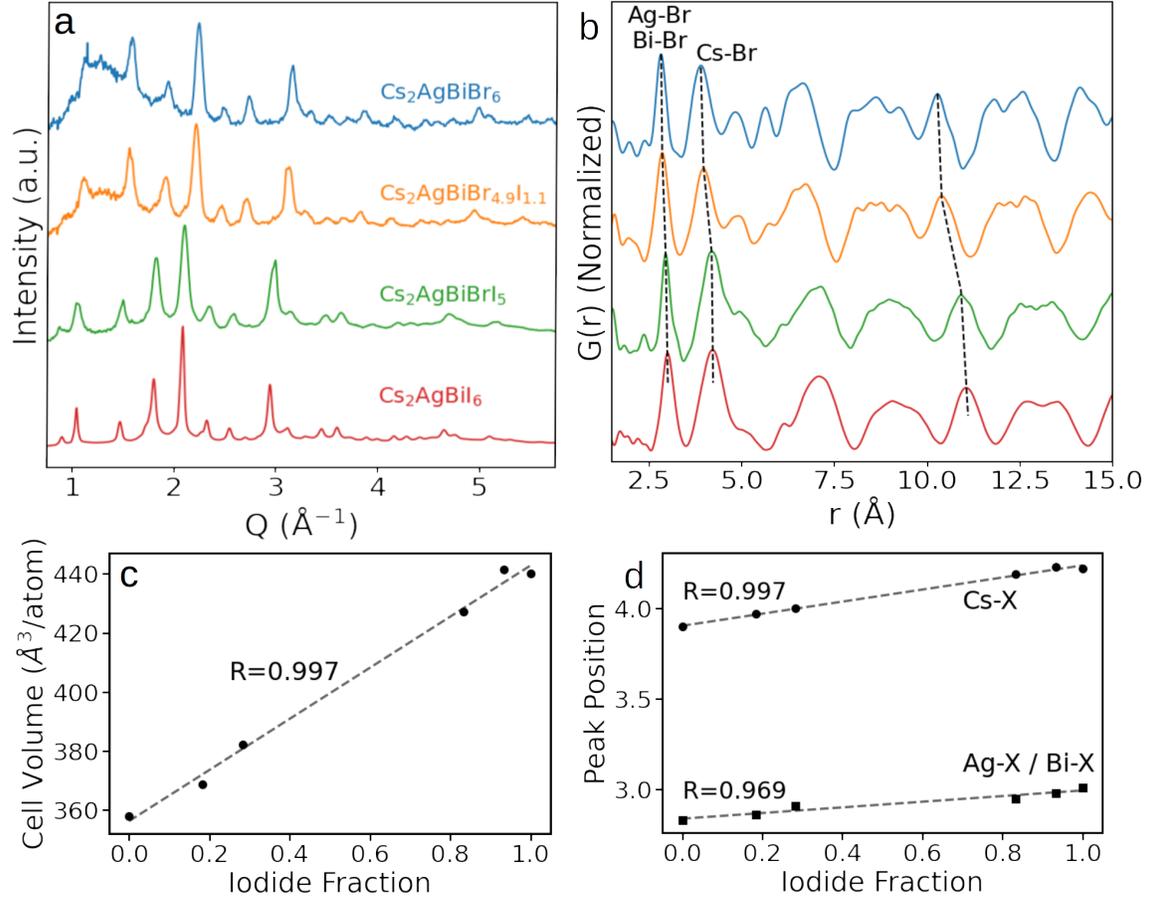

**Figure 3. (a)** High-energy XRD data for select $Cs_2AgBi(Br_{1-x}I_x)_6$ nanocrystal samples from a stepwise anion-exchange reaction series, plotted in Q-space. The first two XRD curves show more noise and background because of weaker signals due to use of less sample in the measurement. **(b)** Atomic Pair Distribution Functions (PDFs, G(r)) for the same nanocrystal series, calculated from X-ray total scattering data. Specific peaks in the short-range order of $Cs_2AgBiBr_6$ are labelled. Dashed lines trace peak positions across multiple samples for these two short-range order peaks as well as for one medium-range order peak. Fourier noise shows up as minor ripples in the PDF data. **(c)** Formula unit volumes obtained from Rietveld refinements of the full series of anion-exchanged samples, plotted as a function of fractional iodide content $x$ in $Cs_2AgBi(Br_{1-x}I_x)_6$, where x is determined independently by EDX measurements. **(d)** Positions of the first two peaks (Ag-X/Bi-X and Cs-X) in the PDF data of panel (b), plotted as a function of fractional iodide content $x$. R values and linear best fits are shown for linear regressions to each peak.

**B. Stability of the iodide elpasolite structure.** After establishing that these iodide NCs form the tetragonal I4-m elpasolite structure, we turn to the question of phase stability, in view of the absence of any other structurally characterized iodide elpasolites and the theoretical predictions that these structures should spontaneously decompose into more stable phases.[5,7,11,13] We may consider two general hypotheses: (1) this iodide elpasolite structure is kinetically trapped, perhaps as a result of anion exchange from a parent bromide elpasolite, and (2) this iodide elpasolite structure is thermodynamically stable, but only because of its surface free energy when prepared at the nanoscale. On one hand, our unsuccessful attempts to prepare $Cs_2AgBiI_6$ NCs directly *via* hot-injection synthesis using TMSI[9] that leave anion exchange from the corresponding bromide NCs as the only synthesis route could suggest that the $Cs_2AgBiI_6$ composition is kinetically trapped. On the other hand, it is difficult to reconcile the notion of a kinetic barrier to ion reorganization with the observation of high ion mobility in these materials during anion exchange. This consideration could suggest instead that the iodide elpasolite NCs are thermodynamically favorable but their direct synthesis is impeded by other kinetically competitive processes, mainly $Cs_3Bi_2I_9$ formation.[9]

Stabilization of metastable polymorphs at the nanoscale is well known,[43-45] generally resulting from increased importance of surface free energy, and $Cs_2AgBiI_6$ may be a new example of this phenomenon. Investigating the possibility that surface energy stabilizes these $Cs_2AgBiI_6$ NCs, we considered two potential sources: structural relaxation at or near the crystal surfaces, and reduction in interfacial free energy due to ligand adsorption. We therefore sought to test both of these hypotheses. To investigate the first, we performed fits to the PDF data for both the tetragonal and cubic structure models (see SI, Fig. S8, S9) and examined these for evidence of structural relaxation. These fits show the greatest differences in the short-range order, consistent with greater deviation from periodicity at the crystal surfaces (which contribute disproportionately to the short-range order for 14 nm NCs), hence suggesting lattice relaxation associated with the crystallite surfaces. These data thus suggest that surface relaxation may play an important role in stabilizing the elpasolite phase on the nanoscale, despite the expected thermodynamic instability of this phase in bulk.[13]

If reduction in interfacial free energy due to ligand adsorption enhances the stability of these iodide elpasolite NCs, then their stability might be expected to depend on the types and quantities of surface ligands present. Although prior work has specifically addressed the influence of various potential surface ligands on the degradation of $CsAgMX_6$ chloride and bromide elpasolite NCs (including $Cs_2AgBiBr_6$),[46,47] such experiments on *iodide* elpasolite NCs have not been reported. In refs 46 and 47, excess primary and tertiary amines were found to be particularly potent for causing complete dissolution of the $CsAgMX_6$ (X = Cl, Br) NCs, attributed to Ag diffusion and the formation of $Cs_3Bi_2Br_9$ and $Cs_3BiBr_6$ phases,[46] or to $M^{3+}$ extraction and formation of $Cs_2AgCl_3$.[47]

To investigate the influence of surface ligation on the stability of the *iodide* elpasolite NCs described here, we performed a series of small-molecule addition experiments in conjunction with the anion-exchange reactions described above, monitored by a combination of XRD, absorption spectroscopy, and TEM. For reference, Fig. 4 highlights key contrasts between the absorption spectra and XRD patterns of $Cs_2AgBiI_6$ and its primary decomposition product, $Cs_3Bi_2I_9$. The $Cs_2AgBiI_6$ absorption spectrum is characterized by a broad onset with shoulders at 390 and 530 nm, whereas $Cs_3Bi_2I_9$ shows more pronounced peaks at 356 and 481 nm along with a shoulder around 420 nm. As noted previously,[9] $Cs_2AgBiI_6$ and $Cs_3Bi_2I_9$ XRD patterns are very similar, due to their similar unit cells, but a signature peak at 27.5° can be used to identify the presence of $Cs_3Bi_2I_9$. Such contrasts help to distinguish between these two compounds and to track any NC decomposition.

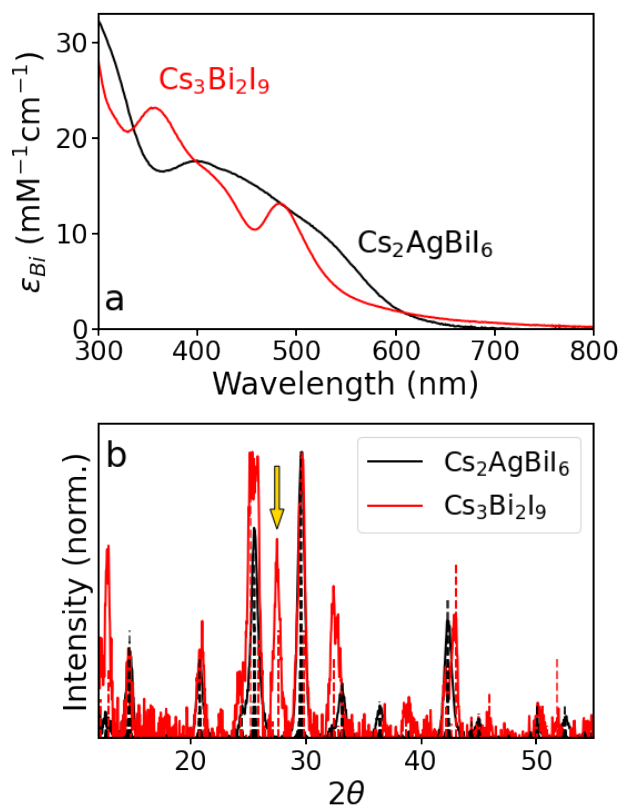

**Figure 4.** Comparison of **(a)** absorption spectra (with molar extinction coefficients referenced to $Bi^{3+}$) and **(b)** XRD patterns for $Cs_2AgBiI_6$ and $Cs_3Bi_2I_9$ nanocrystals, both made *via* hot-injection. The absorption spectra show clear differences in peak positions and shapes between the two compositions. Note that there is some uncertainty in the precise extinction coefficients due to potential error from determination of the $Bi^{3+}$ concentrations. The arrow in (b) highlights a major reflection present in $Cs_3Bi_2I_9$ that is lacking in $Cs_2AgBiI_6$. These contrasts can be used to distinguish between these two phases.

Figure 5 shows XRD data for representative anion-exchange and small-molecule addition reactions, and additional data are provided in the SI. Figure 5a shows results from anion exchange using TMSI (*i.e.*, $Cs_2AgBiBr_6$ + TMSI → $Cs_2AgBiI_6$ + TMSBr)[9,30] followed by addition to the NC solution of one of a set of small molecules that included OA, OLA, sulfobetaine, DDDMABr, and benzyl alcohol. Our NC synthesis was performed in the presence of a mixture of OA and OLA, and further additions of each of these native ligands were thus tested. Sulfobetaine was chosen as a zwitterionic ligand based on its effectiveness in binding to lead-halide perovskite NC surfaces.[48] DDDMABr was selected as an example of a quaternary ammonium species that binds strongly to the surfaces of $CsPbBr_3$ NCs.[49] Benzyl alcohol has been reported to influence ligand binding in perovskite NCs,[50] and may remove surface ligands.[51] Each of these compounds was added to ~1 μM NC solutions at concentrations ranging from ~1 to ~100 μM. For comparison, Fig. 5b shows results from inverting the sequence of additions, *i.e.*, small-molecule addition to a $Cs_2AgBiBr_6$ NC solution, followed by anion exchange. From these data, we observe that the $Cs_2AgBiBr_6$ NCs retain their elpasolite structure upon exposure to all of the compounds tested here, whereas the $Cs_2AgBiI_6$ NCs are far more sensitive, retaining the elpasolite phase with OA or sulfobetaine but transforming to other phases for all other additives examined. The data for the $Cs_2AgBiBr_6$ NCs did show small impurity peaks after exposure to OLA (Fig. S14), but these NCs were substantially more stable than the $Cs_2AgBiI_6$ NCs against degradation by this ligand even at high OLA concentrations, and the dominant phase remained the elpasolite. Additionally, anion exchange in the presence of 50:50 (molar) OA:OLA yielded AgI as the primary product (with evidence of other unidentified minority phases), compared to $Cs_2AgBiI_6$ when just OA was present. Further details of these results are provided as Supporting Information.

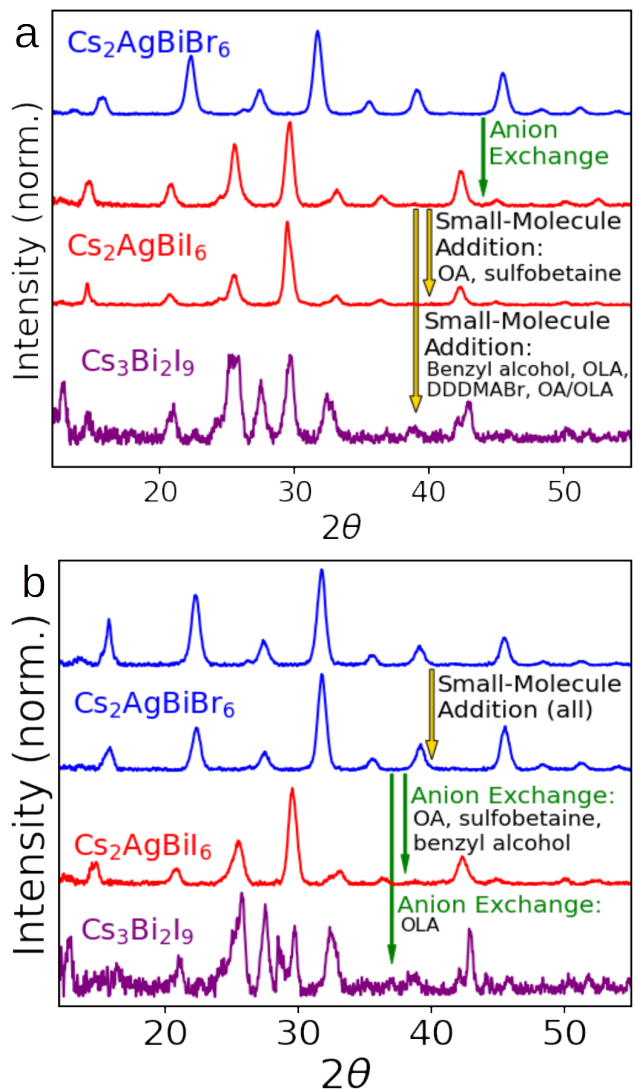

**Figure 5.** Representative XRD data describing the results of anion-exchange and small-molecule addition reactions involving $Cs_2AgBiX_6$ (X = Br, I) nanocrystals. **(a)** Anion exchange followed by small-molecule addition. Addition of oleic acid (OA, data shown here) or sulfobetaine to $Cs_2AgBiI_6$ nanocrystals retains the elpasolite structure, but addition of oleylamine (OLA), 50:50 (mol) OA:OLA, benzyl alcohol (data shown here), or didodecyl-dimethylammonium bromide (DDDMABr) transforms the $Cs_2AgBiI_6$ elpasolite nanocrystals into $Cs_3Bi_2I_9$ nanocrystals. **(b)** Small-molecule additions followed by anion exchange. $Cs_2AgBiBr_6$ nanocrystals retain the elpasolite structure following additions of all small molecules investigated here. Anion exchange for samples with added OA, sulfobetaine, or benzyl alcohol (data shown here) yields elpasolite $Cs_2AgBiI_6$ nanocrystals. Anion exchange for samples with added OLA yields $Cs_3Bi_2I_9$ nanocrystals.

Scheme 1 summarizes the full set of results from these experiments graphically. The arrows denote the reactions that were performed and connect the starting and isolated product compositions for each reaction. The small-molecule additives present during each reaction are shown next to the various arrows. This scheme highlights the different susceptibilities of $Cs_2AgBiBr_6$ and $Cs_2AgBiI_6$ NCs to decomposition. Beginning with the parent $Cs_2AgBiBr_6$ NCs, neither anion exchange nor small-molecule addition causes substantial degradation. From the resulting NCs, however, further chemistry results in a multitude of degradation products. Although $Cs_2AgBiBr_6$ retains its phase upon exposure to the additives examined here, only oleic acid and sulfobetaine preserve the elpasolite structure of $Cs_2AgBiI_6$. Note the pathway-dependent chemical transformations in the case of benzyl alcohol: its addition to $Cs_2AgBiBr_6$ NCs followed by anion exchange yields $Cs_2AgBiI_6$ NCs, whereas its addition to $Cs_2AgBiI_6$ NCs after anion exchange yields a $Cs_3Bi_2I_9$ decomposition product. Additionally, TEM shows the retention of NC size for all reactions, with a morphology change to hexagonal crystals for the hexagonal $Cs_3Bi_2I_9$ decomposition product. We note an apparent correlation between anionic ligands (oleate and sulfobetaine) and stabilities of $Cs_2AgBiI_6$ NCs. It is conceivable that the anionic ligands stabilize the iodide elpasolite NCs by passivating surface anion vacancies. Alternatively, although elpasolite decomposition appears to retain the NC size, we cannot neglect the possibility that some of these additives actively drive decomposition through ion-selective microsolvation. Further investigation will be necessary to fully understand the surface chemistry of this new family of nanomaterials. Nonetheless, we conclude from these results that the stability of elpasolite $Cs_2AgBiI_6$ NCs is indeed highly sensitive to their surface chemistry, and a high surface-to-volume ratio alone is not sufficient to stabilize the elpasolite phase.

**Scheme 1. Summary of Anion-Exchange and Various Small-Molecule Addition Reactions of $Cs_2AgBiX_6$ (X = Br, I) Nanocrystals**

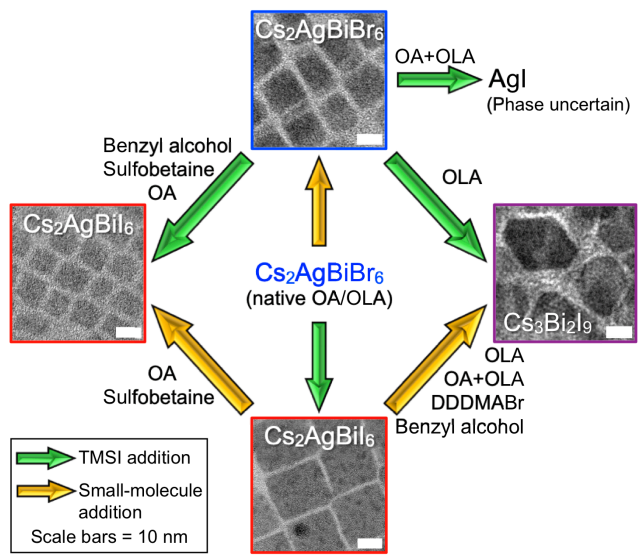

In a separate line of interrogation, we then performed a series of experiments aimed at probing the dependence of $Cs_2AgBiI_6$ stability on temperature and NC size. Thermal stability was probed by both *ex situ* and *in situ* measurements. For the *ex situ* experiments, $Cs_2AgBiI_6$ NCs were deposited onto Si substrates and their structures were monitored as a function of anaerobic anneal temperature and time using XRD and SEM (Fig. S18, S19). The $Cs_2AgBiI_6$ NCs were stable up to ~100°C, above which they began to decompose into a mixture of $Cs_3Bi_2I_9$, CsI, and AgI, with $Cs_3Bi_2I_9$ dominating the XRD patterns (Fig. S18). Full decomposition was observed by 125°C. Moreover, SEM of the $Cs_3Bi_2I_9$ product (Fig. S19) revealed extensive particle sintering. Heating for 2 h at temperatures below 100°C caused no decomposition. These results demonstrate that the decomposition of $Cs_2AgBiI_6$ NCs is thermally activated. Very similar results are obtained from *in situ* XRD measurements during heating in air (Fig. S20).

From the above observations, we hypothesized that decomposition may be linked to the reduced surface-to-volume ratios in the larger crystallites that form upon sintering, *i.e.*, the reduced influence of surface free energy. To test this hypothesis, we prepared samples that deliberately contained broad size distributions. For example, Fig. 6 shows XRD, TEM, and electron diffraction (ED) data for a sample containing a mixed population of nanocubes (L ~ 15 nm) and nanorods (L ~ 200 nm, W ~ 15 nm). TEM shows that nanorods are the dominant species in this sample, but critically, XRD confirms the presence of *only* the elpasolite phase (Fig. 6a), contrary to the above hypothesis. Electron diffraction (Fig. 6c) from the area imaged in Fig. 6b shows signature arcs (denoted with red arrows) corresponding to preferential diffraction from the one-dimensional nanorods, associated with net spatial orientation of the nanorods in Fig. 6b. These arcs overlay the elpasolite diffraction rings, confirming that these nanorods share the iodide elpasolite structure. There is thus no evidence for a correlation between particle size and phase transformation in this size regime.

We then hypothesized that decomposition may be associated with the combination of heat and particle size. To test this hypothesis, we performed *in situ* heating during TEM measurements of another sample possessing a deliberately broad size distribution (Fig. S21, S22). Again, we found no correlation between particle size and decomposition, nor did we generally observe particle sintering upon heating. Some particles transformed and sintered at 100°C while others of the same size persisted as elpasolites even to 120°C (Fig. S21). Additionally, other NCs decomposed without changing size or sintering with their neighbors. We conclude that phase transformation and particle sintering are *not* linked and instead can occur independently upon heating. Instead, the results suggest that the critical factor in the thermally induced phase transformation of $Cs_2AgBiI_6$ is likely the loss of surface ligands. In support of this conclusion, we have found that colloidal $Cs_2AgBiI_6$ NCs retain the $Cs_2AgBiI_6$ phase even when heated in solution to 140°C (Fig. S23), beyond the ~100-125°C decomposition temperature of the same NCs on a solid substrate. Collectively, these results provide strong support for the conclusion that surface ligands are a primary factor determining the stability of the $Cs_2AgBiI_6$ elpasolite phase and preventing its transformation to $Cs_3Bi_2I_9$, and that reduced dimensionality is necessary but not sufficient. Given these findings, we hypothesize that judicious choice or exclusion of surface ligands, possibly in conjunction with other crystallite size and surface modifications (*e.g.*, surface cation exchange), may ultimately allow optimization of the surface free energies to achieve maximum stabilization of the iodide elpasolite phase relative to its decomposition products.

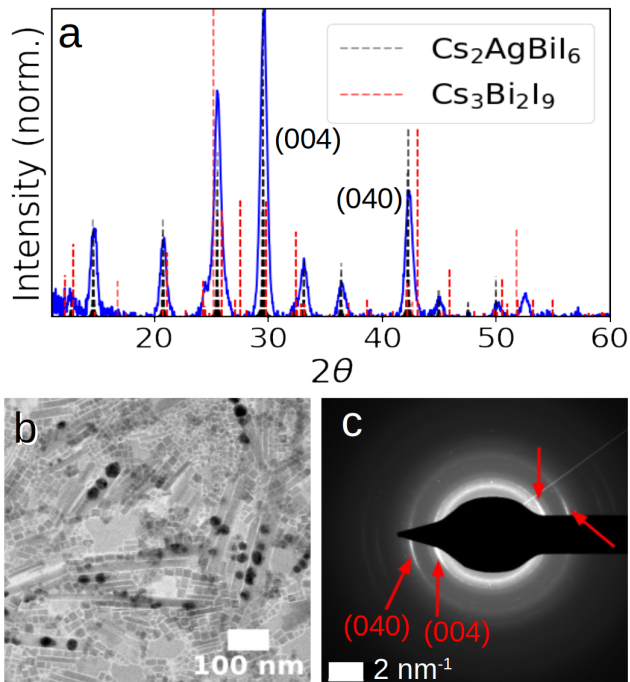

**Figure 6. (a)** XRD, **(b)** TEM, and **(c)** electron diffraction (ED) of a $Cs_2AgBiI_6$ sample with a mixed population of nanocubes and 1D nanorods. The red arrows in (c) identify diffraction arcs corresponding to the nanorods. Note the net spatial alignment of the rods in panel (b), giving rise to the arcs in panel (c).

Finally, to test the role of purely kinetic trapping in the formation and stability of this iodide elpasolite, we prepared $Cs_2AgBiBr_6$ polycrystalline thin films (100-200 nm grain sizes, see SI Fig. S17) by thermal evaporation and attempted to convert these to $Cs_2AgBiI_6$ by anion exchange, as was done with the NCs. Notably, these grains possess no surface ligands and have much lower surface-to-volume ratios than the colloidal NCs (~0.04 nm$^{-1}$ *vs* ~0.40 nm$^{-1}$), but ion mobility is sufficient for the entire grain volume to remain accessible (see, *e.g.*, refs. [30,52,53] for analogous anion-exchange reactions on lead-halide perovskite films). We previously demonstrated a strong thermodynamic driving force and near-stoichiometric reactivity for TMSX reagents in anion exchange of both perovskites and elpasolites.[9,30] Figure 7 presents absorption and XRD data for a representative anion-exchange experiment on one of these thin films. With a single dose (5 molar equivalents) of gas-phase TMSI, the absorption spectrum shows the pronounced low-energy peak decrease in intensity and shift to slightly longer wavelengths, consistent with partial anion exchange. With a second dose of TMSI, this absorption band simply decreases in intensity and the optical quality of the film was observed by eye to degrade, suggesting decomposition. The XRD data in Fig. 7b confirm these observations, showing a shift in the elpasolite peak positions with the first dose of TMSI corresponding to anion alloying and formation of $Cs_2AgBi(Br_{0.66}I_{0.34})_6$, but no further movement of the elpasolite peaks is observed upon delivering the second dose of TMSI. Instead, peak broadening is observed, consistent with a decrease in crystallinity. Anion exchange proceeded more slowly in these thin films than in the NCs, and we therefore allowed samples to react for ~24 h to ensure completion after each dose. Additional doses of TMSI or slower dosing with longer reaction times (Fig. 7c, S16) both failed to yield any further anion exchange beyond $Cs_2AgBi(Br_{0.66}I_{0.34})_6$. Using an extremely large dose of TMSI (~100-fold excess) decomposed the sample to yield a mixture of $Cs_3Bi_2I_9$, AgI, and CsI, with additional minor impurities. Overall these results indicate that the elpasolite structure is stable only up to $Cs_2AgBi(Br_{0.66}I_{0.34})_6$ in these films, and further conversion is not possible. This result provides evidence for rejection of the hypothesis that the iodide elpasolite NCs are achieved solely due to kinetic trapping, and it is consistent with the conclusion drawn above that surface free energy (*i.e.*, thermodynamics) plays a critical role in stabilizing the iodide elpasolite phase.

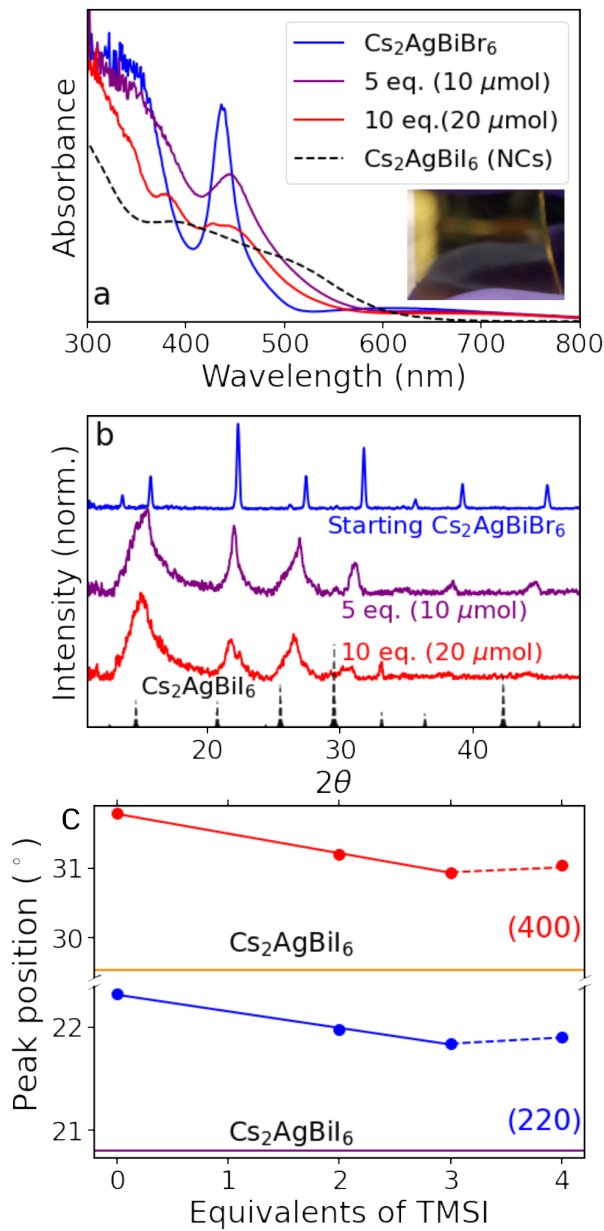

**Figure 7. (a)** Absorption spectra (inset: photograph of starting thin film) and **(b)** XRD data collected for a thermally evaporated thin film of $Cs_2AgBiBr_6$ at various stages of TMSI addition. A stoichiometric excess of TMSI (10 µmol, similar to the quantities used in the above nanocrystal reactions) was used in both additions and allowed to react for ~24 h. Analysis of the XRD peak shifts suggests a terminal composition of $Cs_2AgBi(Br_{0.66}I_{0.34})_6$. **(c)** Peak positions as a function of equivalents of TMSI. Further anion exchange was not observed beyond 3 equivalents. For reference, the yellow and purple horizontal lines denote the positions of the (400) and (220) peaks in $Cs_2AgBiI_6$, respectively.

One possible balanced chemical reaction describing $Cs_2AgBiI_6$ decomposition is given by eq. 1.

$$2Cs_2AgBiI_6 \rightarrow Cs_3Bi_2I_9 + 2AgI + CsI \qquad (1)$$

The free-energy driving force for this reaction at the nanoscale ($\Delta G_{rxn,NC}$) involves contributions from the lattice (taken as $\Delta G_{rxn,bulk}$), from the change in ligand binding between reactants and products ($\Delta G_{ligand}$), and from any change in intrinsic surface free energies between reactants and products ($\Delta G_{surface}$). The data presented here argue that the magnitude of $\Delta G_{ligand} + \Delta G_{surface}$ must exceed that of $\Delta G_{rxn,bulk}$ for at least some of the ligands examined here. A quantitative analysis of these free energies is complicated by the fact that the thermodynamics of ligand binding to $Cs_2AgBiI_6$, $Cs_3Bi_2I_9$, AgI, and CsI NCs are not documented. Additionally, decomposition does not generally yield the stoichiometric ratio of crystalline products described by eq. 1 (*e.g.*, Scheme 1), indicating that additional product species must be accounted for. A full thermodynamic analysis of $Cs_2AgBiI_6$ NC decomposition will thus require careful experimentation to quantify these many contributing factors.

**Conclusion**

The data and analysis presented here provide the first definitive structural characterization of any iodide elpasolite. Rietveld refinement of high-energy XRD and PDF data collected from $Cs_2AgBiI_6$ nanocrystals shows them to exhibit an elpasolite structure with tetragonal (I4-m) symmetry. High-energy XRD and PDF measurements show that the elpasolite structure gradually expands during anion exchange to accommodate the larger iodide ions as it evolves continuously from cubic $Cs_2AgBiBr_6$ to tetragonal $Cs_2AgBiI_6$. Fitting the PDF data using the refined structure model reveals evidence of surface relaxation, and additional experiments probing the effects of surface ligation and grain size strongly support the conclusion that the elpasolite structure of $Cs_2AgBiI_6$ is stabilized by the surface free energy (*i.e.*, thermodynamics) of these nanocrystals. The detailed structural characterization presented here, along with the identification of factors influencing stability of the elpasolite phase, provides a valuable foundation for future exploration and application of this and yet-undiscovered members of the extremely rare iodide elpasolite family of compounds. These results further emphasize the unique opportunities that exist for developing this family of compounds when working at the nanoscale.

**Supporting Information**

The Supporting Information, containing additional figures & tables and the cif files for each structure refinement, is available free of charge at [URL to be inserted by the publisher].

**Acknowledgments.** This research was primarily supported by the UW Molecular Engineering Materials Center, an NSF Materials Research Science and Engineering Center (Grant No. DMR-1719797). Synchrotron X-ray data analysis was supported by the DOE Office of Science, Office of Basic Energy Sciences, Chemical Sciences, Geosciences, and Biosciences Division, Geosciences Program at Pacific Northwest National Laboratory (PNNL) under FWP 56674. PNNL is a multiprogram national laboratory operated for the DOE by Battelle Memorial Institute under Contract DE-AC05-76RL0-1830. Part of this work was conducted at the Molecular Analysis Facility, a National Nanotechnology Coordinated Infrastructure (NNCI) site at the University of Washington, which is supported in part by funds from the National Science Foundation (awards NNCI-2025489 and NNCI-1542101), the Molecular Engineering & Sciences Institute, and the Clean Energy Institute. We acknowledge Scott Braswell for his assistance with SEM & EDX measurements. TEM heating data using an *in situ* heating holder were collected in the William R. Wiley Environmental Molecular Sciences Laboratory (EMSL), a national scientific user facility sponsored by DOE's Office of Biological and Environmental Research and located at Pacific Northwest National Laboratory. We thank Libor Kovarik for his assistance with these experiments. This research used resources of the Advanced Photon Source, a U.S. Department of Energy (DOE) Office of Science User Facility operated for the DOE Office of Science by Argonne National Laboratory under Contract No. DE-AC02-06CH11357. The authors thank Leighanne Gallington and Olaf Borkiewicz from Sector 11 (BL 11-ID-B) for assistance collecting data presented in this manuscript.

**Table of Contents Graphic**

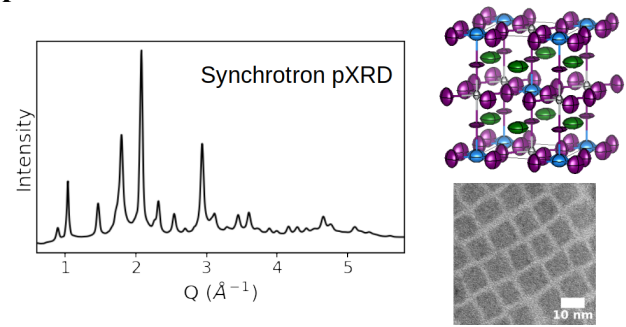

*Supporting Information for*

# Structure and Stability of the Iodide Elpasolite, $Cs_2AgBiI_6$

Kyle T. Kluherz,[1] Sebastian T. Mergelsberg,[2] James J. De Yoreo,[2] and Daniel R. Gamelin[1,*]

[1]*Department of Chemistry, University of Washington, Seattle, WA 98195, U.S.A.*
[2]*Physical Sciences Division, Pacific Northwest National Laboratory,
Richland, WA 99352, U.S.A.*
Email: gamelin@uw.edu

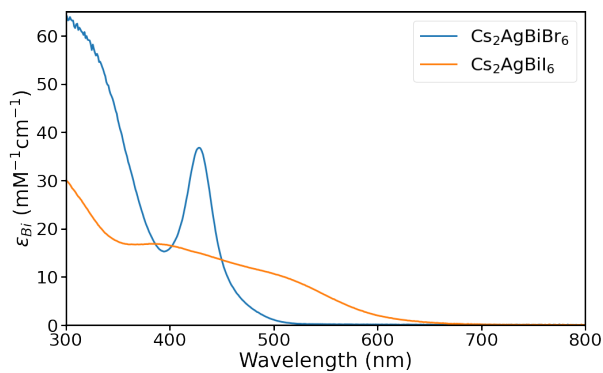

**Figure S1.** Absorbance of $Cs_2AgBiBr_6$ nanocrystals (blue) and anion-exchanged $Cs_2AgBiI_6$ nanocrystals (orange) used for X-ray total scattering measurements. Note that there is some uncertainty in the precise extinction coefficients due to potential error in determination of the $Bi^{3+}$ concentrations.

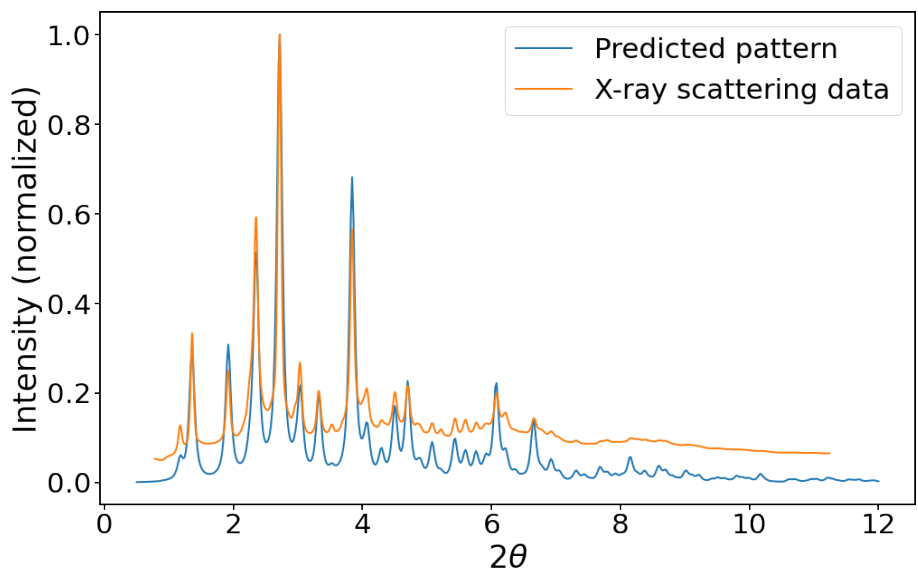

**Figure S2.** X-ray scattering data (orange) collected for $Cs_2AgBiI_6$ nanocrystals in hexanes compared with predicted scattering pattern (blue) from a simple cubic (Fm3m) model made by replacing Br in $Cs_2AgBiBr_6$ with I and expanding the unit cell.[1]

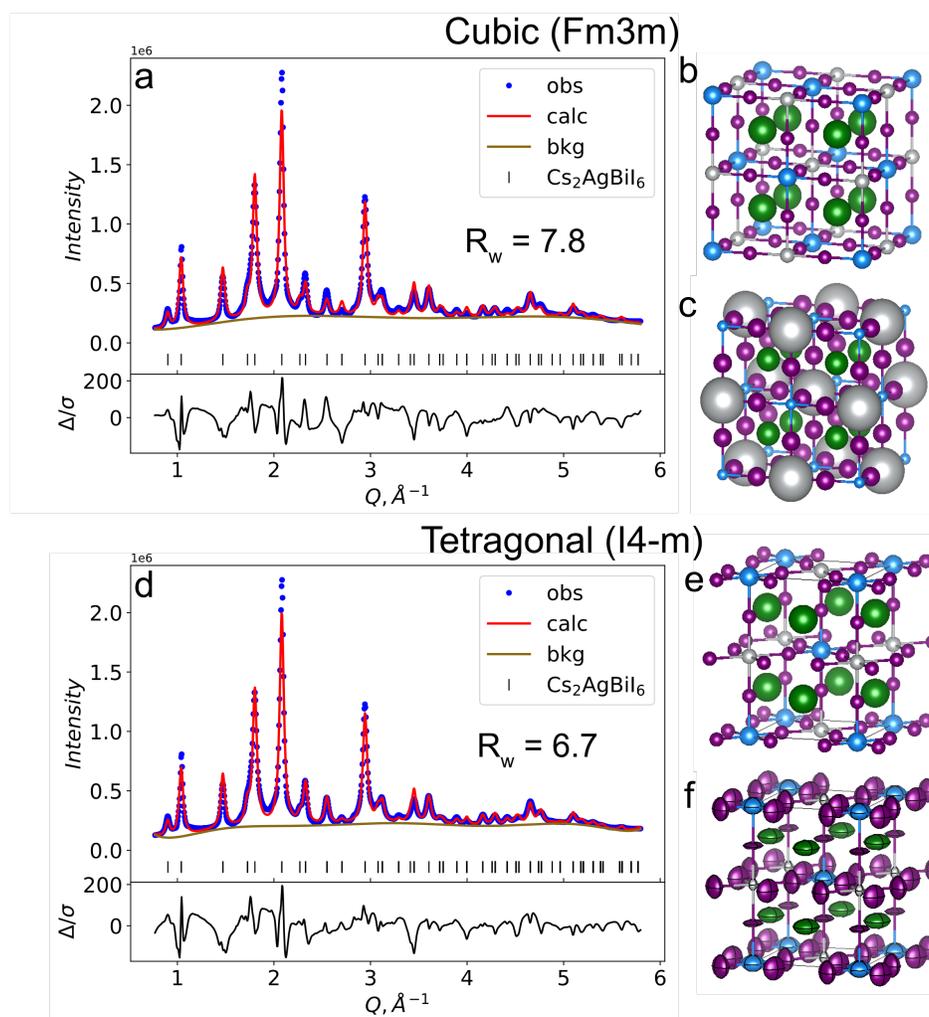

**Figure S3.** Comparison of Rietveld refinement results using synchrotron X-ray scattering data for $Cs_2AgBiI_6$ nanocrystals in hexanes solution. **(a)** Data (blue dots), calculated pattern (red curve), background (brown curve), predicted peak positions (vertical black lines), and residual of fit ($\Delta/\sigma$) obtained using cubic Fm3m structure. $R_w$ is the overall weighted R-factor for the refinement. **(b)** Cubic Fm-3m structure model. Silver: $Ag^+$; blue: $Bi^{3+}$; green: $Cs^+$; purple: $I^-$. **(c)** Structure model produced from (a) showing 95% atomic displacement parameters (ADPs). Note the very large displacements of $Ag^+$ atoms (0.45 Å). **(d)** Data (blue dots), calculated pattern (red curve), background (brown curve), predicted peak positions (vertical black lines), and residual of fit ($\Delta/\sigma$) obtained using tetragonal I4-m structure. **(e)** Tetragonal I4-m structure model. **(f)** Structure model produced from (e) showing 95% ADPs refined using PDF data.

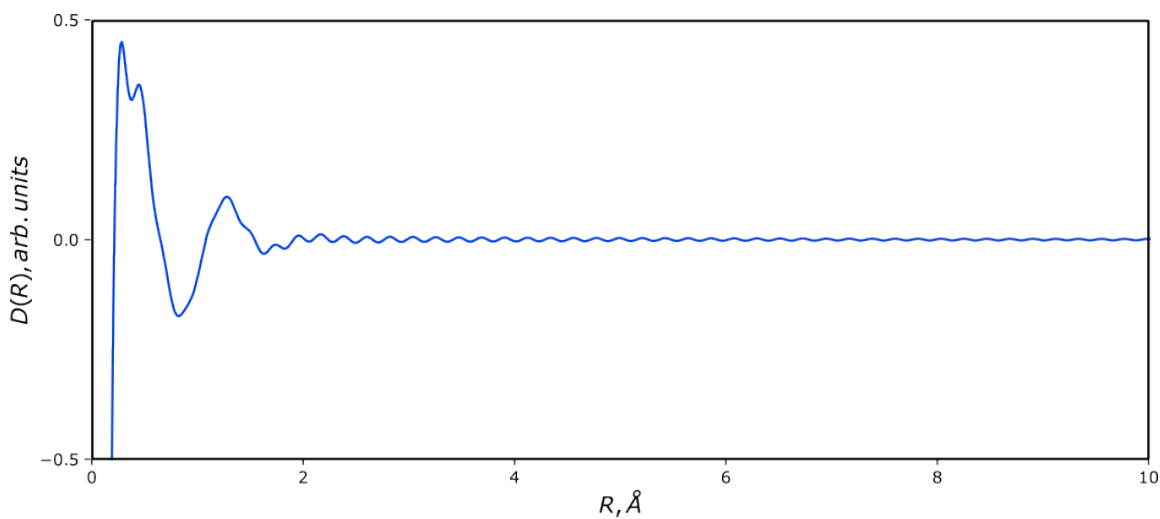

**Figure S4.** Example radial distribution function of the residuals from the Rietveld refinements. The two prominent peaks are below 2 Å, indicating no significant contribution of the elpasolite or secondary crystalline phases. The broad peak between 1 and 2 Å likely corresponds to C−C, C−O, and C−N atom pairs of the ligands and the ligand-solvent interface.

**Table S1.** Residuals after last Rietveld refinement for each structure considered here.

| Structure | $R_w$ | Red $\chi^2$ | | |
|---|---|---|---|---|
| | | 250 mm | 1000 mm | 1500 mm |
| Fm-3m | 7.8 | 2.5478 | 2.1691 | 2.3527 |
| I4-m | 6.7 | 1.7386 | 1.6154 | 1.7935 |
| R-3m | 13.7 | 6.0920 | 6.5109 | 7.7594 |

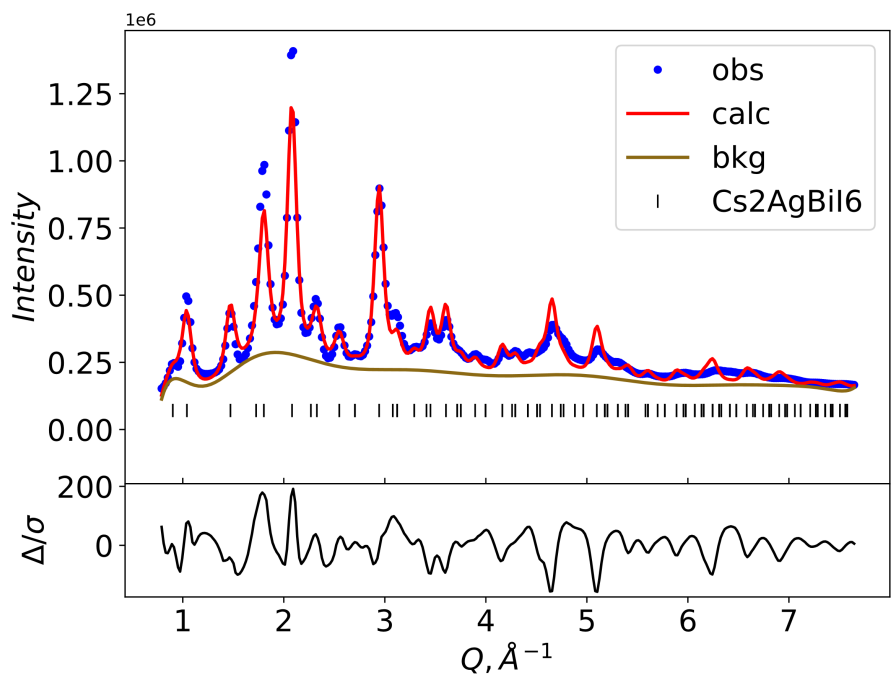

**Figure S5.** Rietveld refinement results using synchrotron X-ray scattering data. Data, calculated pattern, background, predicted peaks, and residual of fit using cubic Fm-3m structure with Ag $U_{iso}$ parameter reset to 0.02 and not refined. Weighted residual $R_w$ = 12.9. Note additional peaks in the residual.

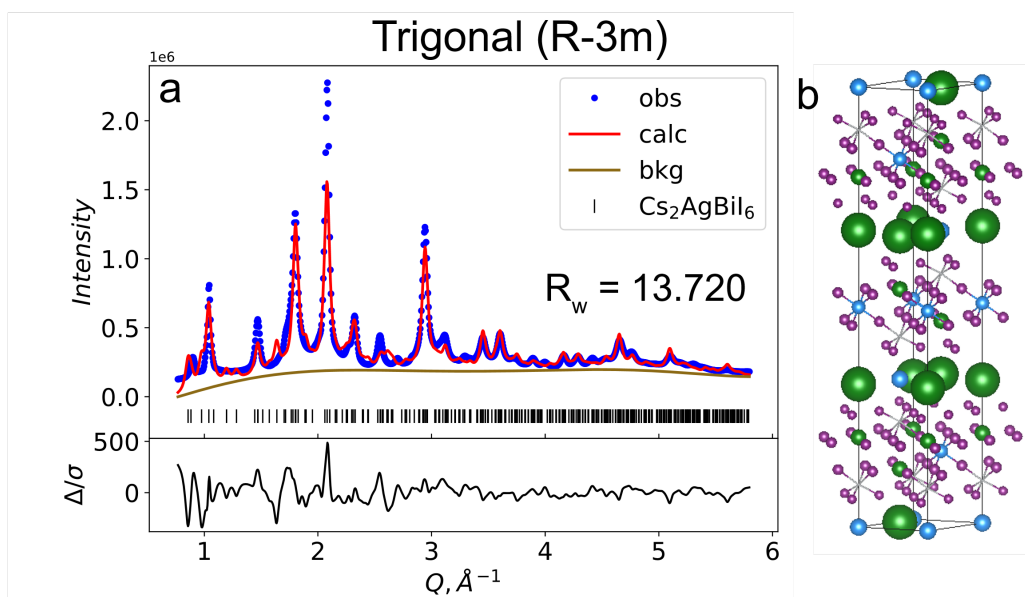

**Figure S6.** Rietveld refinement results using synchrotron X-ray scattering data. **(a)** Data, calculated pattern, background, predicted peaks, and residual of fit using trigonal R/3m structure. Weighted residual $R_w$ = 13.7. **(b)** Trigonal structure model showing 95% atomic displacement parameters (ADPs) from fit. Silver: $Ag^+$; blue: $Bi^{3+}$; green: $Cs^+$; purple: $I^-$.

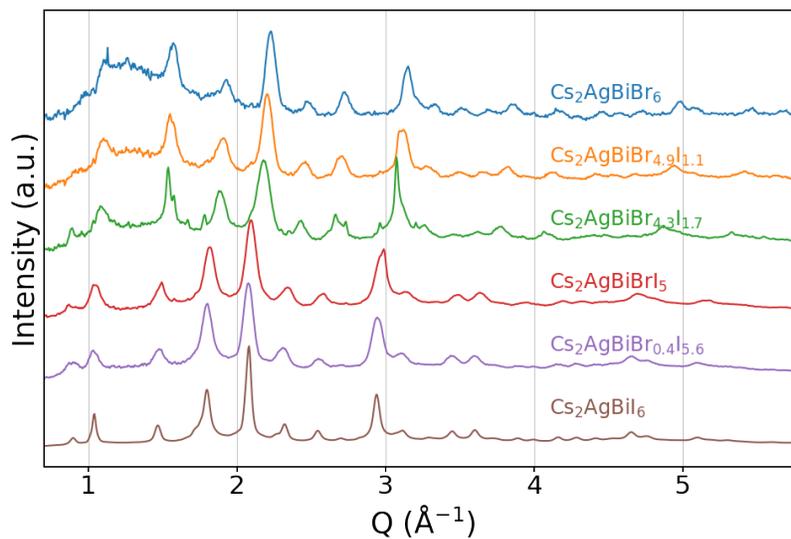

**Figure S7.** Synchrotron XRD patterns for each sample in a series from $Cs_2AgBiI_6$ to $Cs_2AgBiI_6$ plotted in Q-space. Greater noise in the $Cs_2AgBiBr_6$, $Cs_2AgBiBr_{4.3}I_{1.7}$, and $Cs_2AgBiBr_{4.9}I_{1.1}$ patterns is due to the lower signal-to-noise ratio for those samples. Weighted residuals for each fit from $Cs_2AgBiBr_6$ to $Cs_2AgBiBr_{0.4}I_{5.6}$, respectively: $R_w$ = 7.84, 8.32, 9.86, 6.68, 6.06.

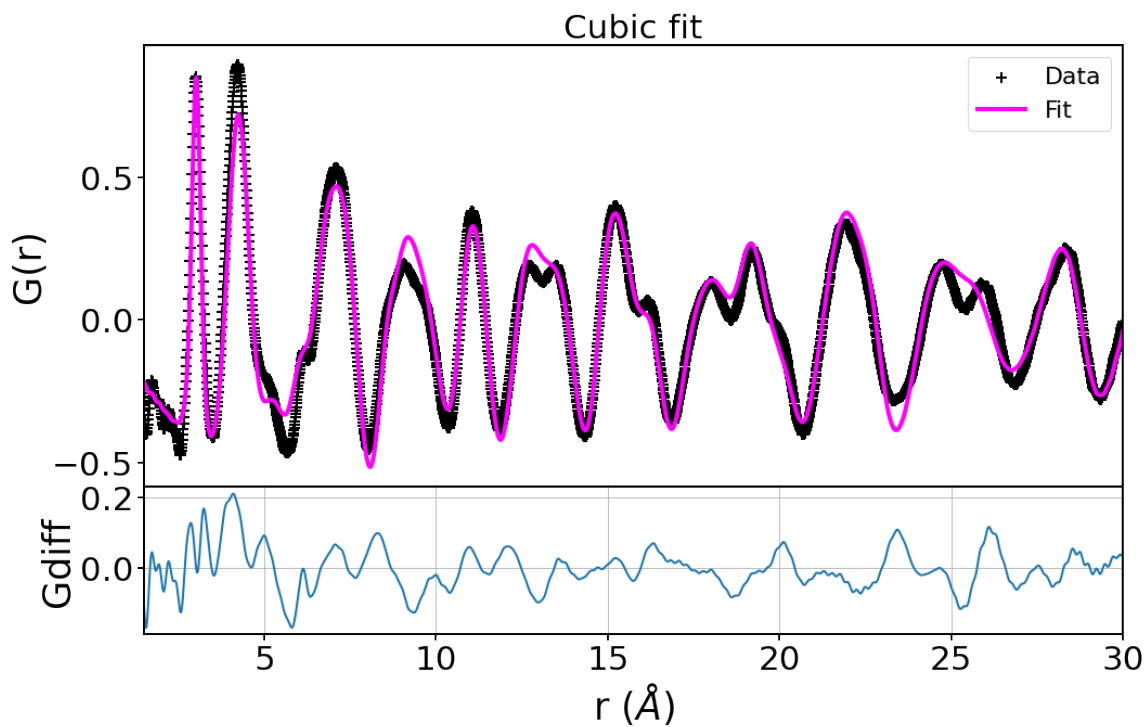

**Figure S8.** Real-space fit to the $Cs_2AgBiI_6$ PDF data using a cubic (Fm-3m) structural model, run using pdfgui. $G_{diff}$ shows the difference between the fit and the data. $R_w$ = 0.227, reduced chi squared = 0.0035.

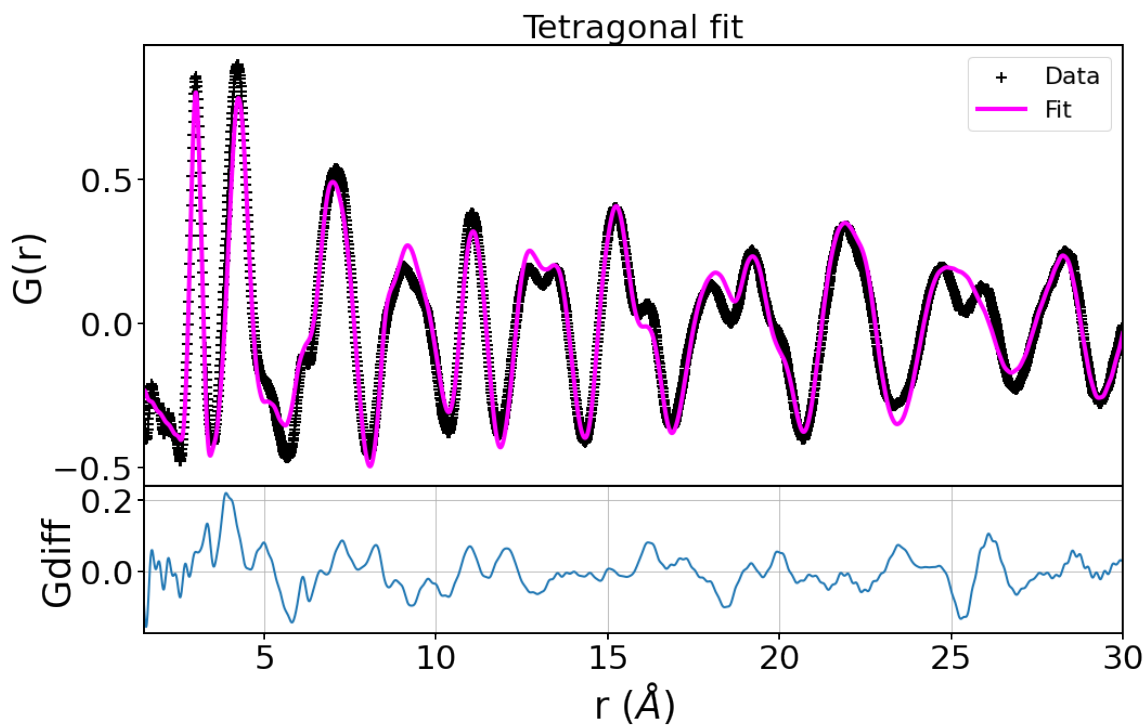

**Figure S9.** Real-space fit to the $Cs_2AgBiI_6$ PDF data using a tetragonal (I4-m) structural model, run using pdfgui. $G_{diff}$ shows the difference between the fit and the data. $R_w$ = 0.207, reduced chi squared = 0.0030.

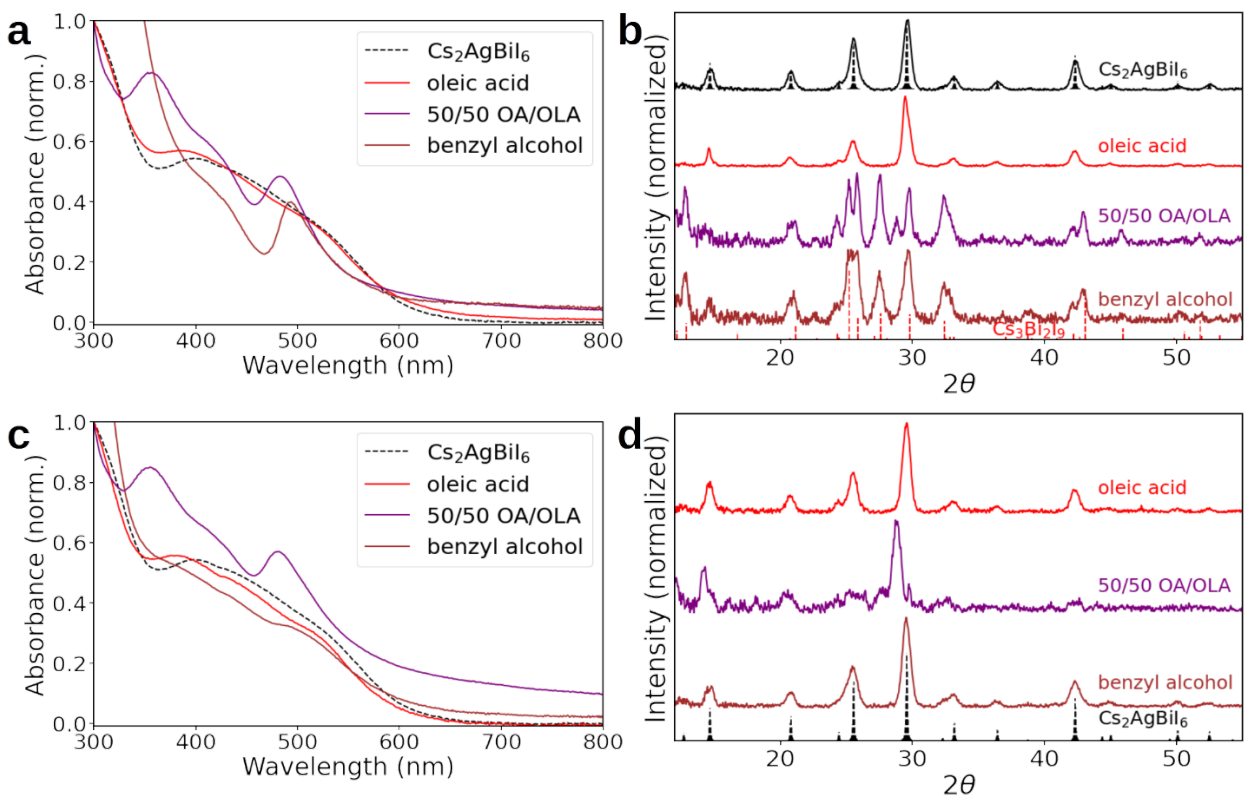

**Figure S10.** Absorption spectra and XRD patterns for $Cs_2AgBiI_6$ nanocrystals exposed to oleic acid, 50:50 (mol) oleic acid (OA) and oleylamine (OLA), and benzyl alcohol. **(a,b)** Data from samples in which $Cs_2AgBiBr_6$ nanocrystals were first converted to $Cs_2AgBiI_6$ by anion exchange, and the resulting $Cs_2AgBiI_6$ nanocrystals were then exposed to the respective ligands. **(c,d)** Data from samples in which $Cs_2AgBiBr_6$ nanocrystals were first exposed to these ligands, and then converted to $Cs_2AgBiI_6$ (or decomposition products) by anion exchange.

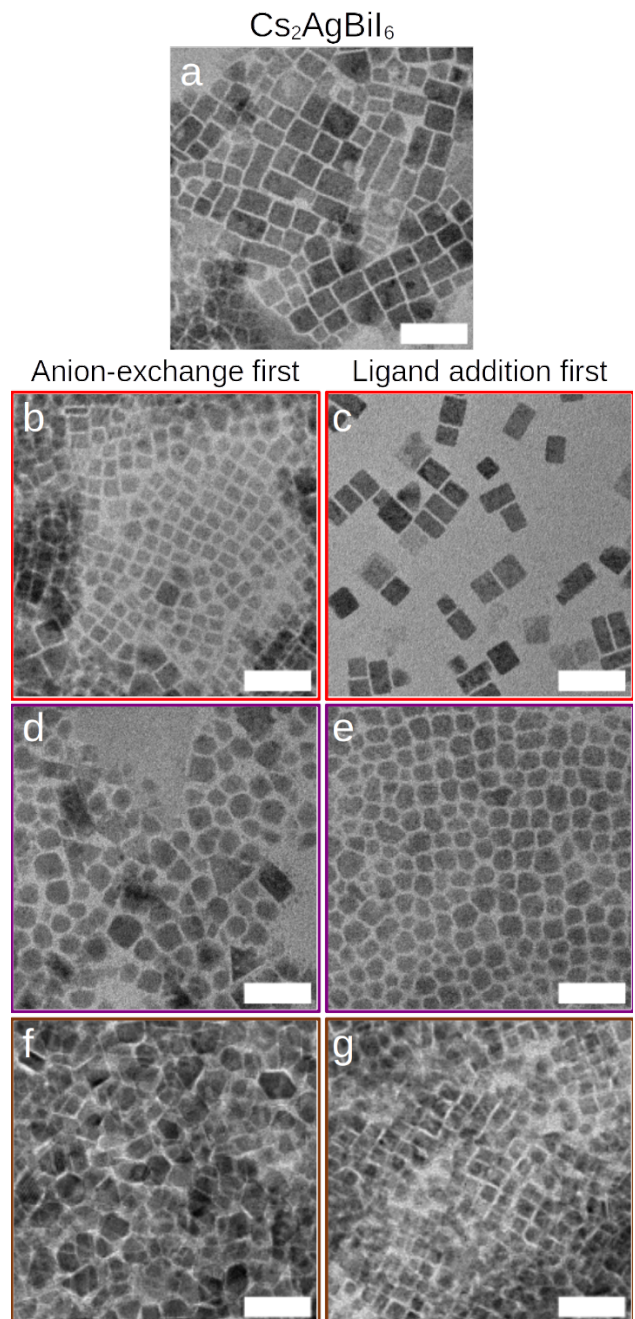

**Figure S11.** TEM images of $Cs_2AgBiX_6$ (X = Br, I) nanocrystals exposed to various organic compounds, as described in above Fig. S10. **(a)** As-made $Cs_2AgBiI_6$ nanocrystals. **(b,d,f)** $Cs_2AgBiI_6$ nanocrystals exposed to organic compounds. **(c,e,g)** $Cs_2AgBiBr_6$ nanocrystals exposed to organics then subsequently converted to $Cs_2AgBiI_6$ (or decomposition products) by anion exchange. Organic compounds added are as follows: (b) and (c) oleic acid; (d) and (e) 50:50 mol % OA:OLA; (f) and (g) benzyl alcohol. Scale bars are 50 nm.

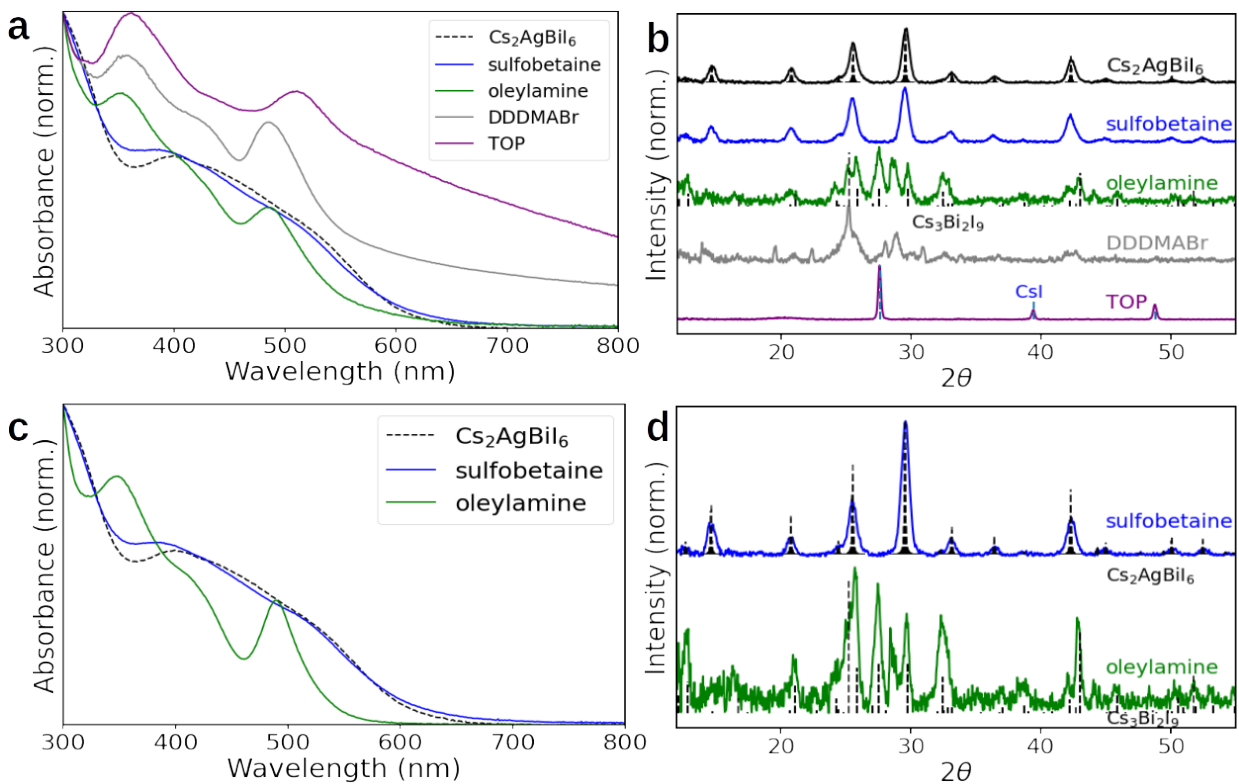

**Figure S12.** Absorption (**a**, **c**) and XRD (**b**, **d**) data for $Cs_2AgBiI_6$ nanocrystals exposed to sulfobetaine, oleylamine, DDDMABr, and TOP. **(a)** and **(b)**: $Cs_2AgBiI_6$ nanocrystals directly exposed to the respective organics. **(c)** and **(d)**: $Cs_2AgBiBr_6$ nanocrystals exposed to organics, then anion-exchanged to $Cs_2AgBiI_6$ using a standard anion-exchange procedure.

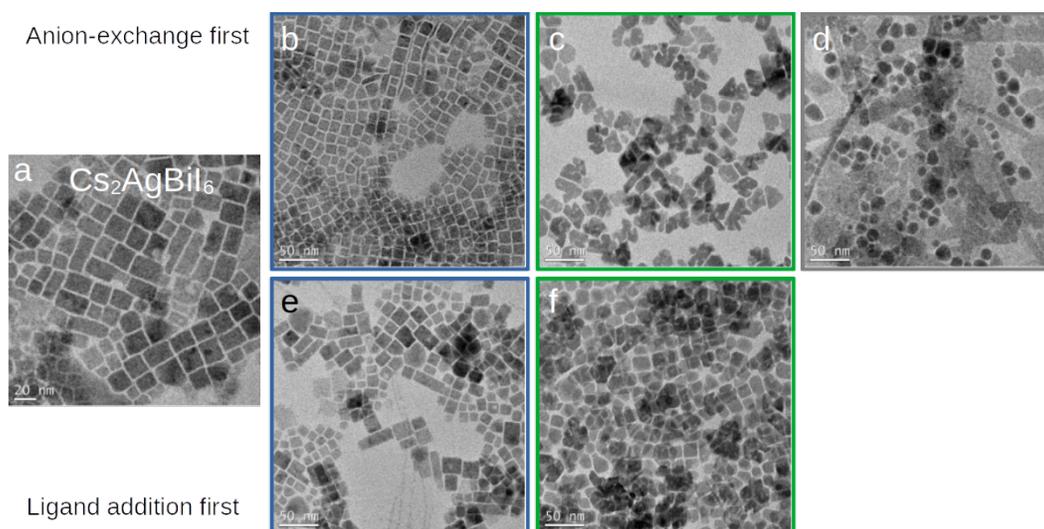

**Figure S13.** TEM images of the $Cs_2AgBiX_6$ (X = Br, I) nanocrystals from Fig. S12. **(a)** As-made $Cs_2AgBiI_6$. Top row **(b,c,d)** $Cs_2AgBiI_6$ nanocrystals exposed to select small molecules. Bottom row **(e,f)** $Cs_2AgBiBr_6$ nanocrystals exposed to small molecules then subsequently anion-exchanged to $Cs_2AgBiI_6$. Small molecules added are as follows: (b) and (e) - sulfobetaine; (c) and (f) - oleylamine; (d) DDDMABr. Scale bars are 50 nm.

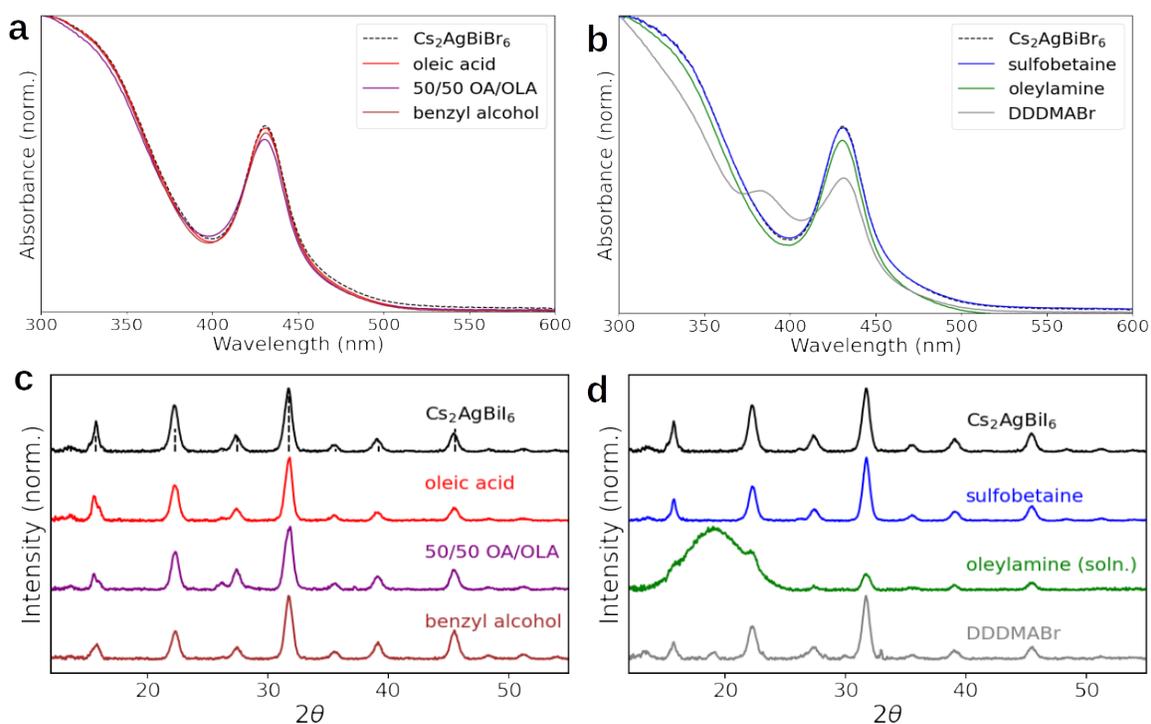

**Figure S14.** Absorption (**a**, **b**) and XRD (**c**, **d**) data for $Cs_2AgBiBr_6$ nanocrystals exposed to various small molecules. **(a/c)** Oleic acid, 50/50 mol/mol oleic acid (OA) and oleylamine (OLA), and benzyl alcohol. **(b/d)** sulfobetaine (3-(N,N-dimethyl-octadecylammonio)-propanesulfonate), oleylamine, and DDDMABr (didodecyl-dimethylammonium bromide). Oleylamine data in (d) were measured in solution for a more accurate comparison with anion exchange conditions. It was found that oleylamine-exposed $Cs_2AgBiBr_6$ nanocrystals were unstable once solvent evaporated.

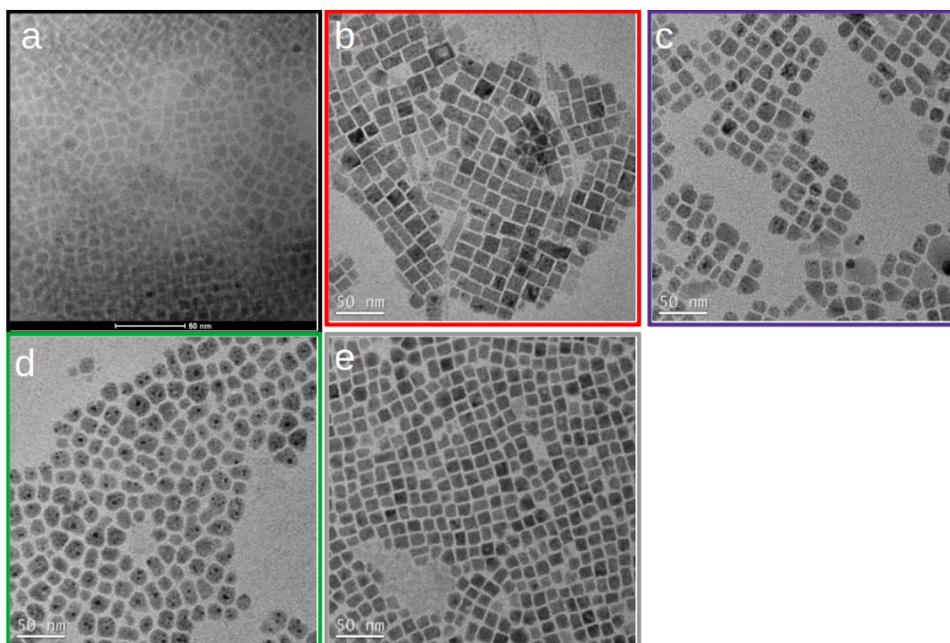

**Figure S15.** TEM images of above $Cs_2AgBiBr_6$ nanocrystals exposed to select small-molecule additives. **(a)** Starting $Cs_2AgBiBr_6$. $Cs_2AgBiBr_6$ exposed to: **(b)** Oleic acid. **(c)** 50/50 OA/OLA. **(d)** OLA. **(e)** DDDMABr.

**Table S2**. Water contents of select additives and solvents measured using Karl-Fischer titration.

| Reagent | Hexanes (dry) | Hexanes | Benzyl alcohol | Oleic Acid | Oleylamine | DDDMABr |
|---|---|---|---|---|---|---|
| Water content (ppm) | 0.0 | 58.3 | 86 | 74.2 | 86.1 | 13.4 |

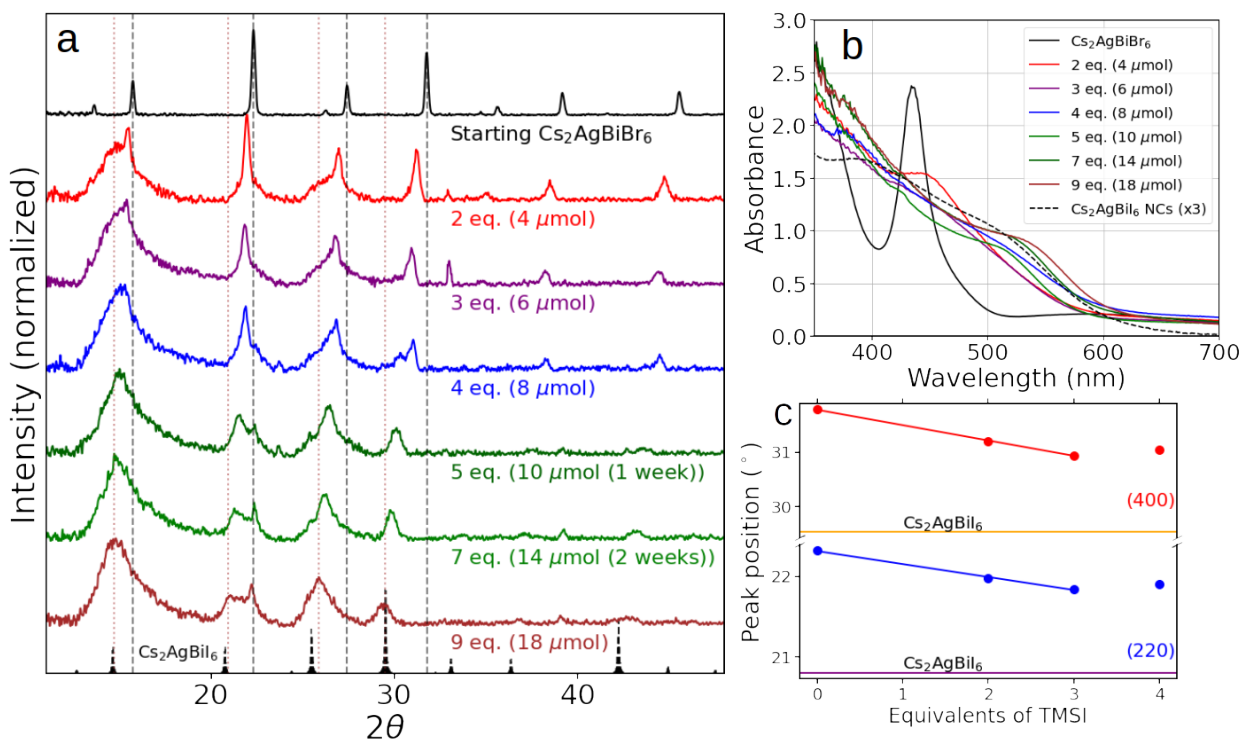

**Figure S16.** Gradual anion exchange of a thermally evaporated thin film of $Cs_2AgBiBr_6$ using TMSI vapors. **(a)** XRD patterns of thin film after each TMSI exposure, given in molar equivalents and μmol quantities. Unless otherwise specified, films were allowed to react for 24 h before stopping the reaction. **(b)** Absorption spectra of each thin film exposed to identical conditions to those shown in **(a)**. **(c)** Scatter plot of peak positions of the (400) and (220) peaks in (a) for the first 4 films. A linear trend was observed for the first 2 TMSI additions. After 4 equivalents, it becomes unclear whether the film can still be considered an elpasolite, thus the peaks were not identified.

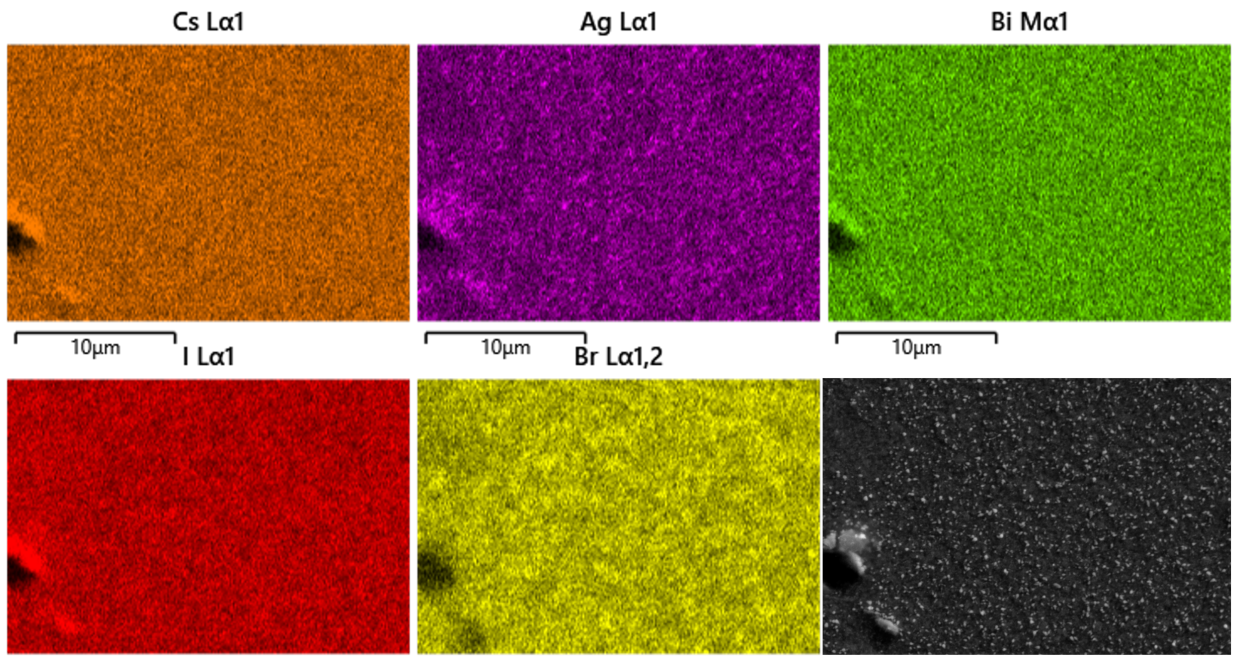

**Figure S17**. SEM EDX mapping of final exposure Cs$_2$AgBiBr$_6$ polycrystalline thin film from slow anion-exchange reaction series. Maps for I, Cs, Br, Bi, and Ag are shown, along with the base SEM image. Grain sizes were observed to be in the range of 100-200 nm.

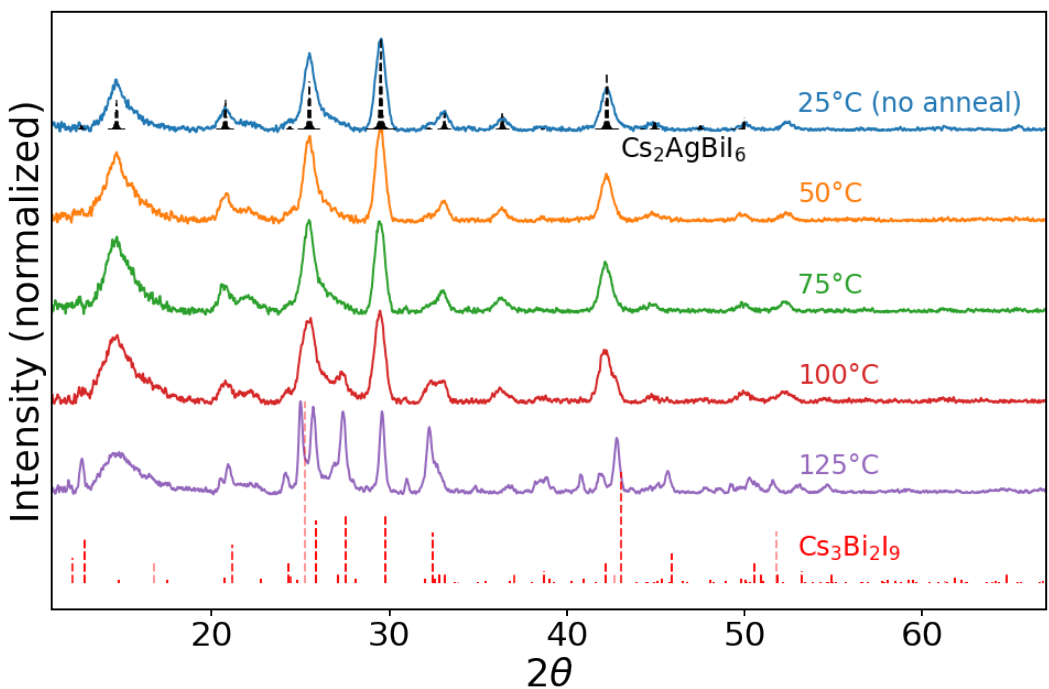

**Figure S18.** XRD patterns of Cs$_2$AgBiI$_6$ nanocrystal film samples heated to various temperatures under nitrogen atmosphere for 30 min (*ex situ*). Measurements were collected in air promptly after cooling the samples.

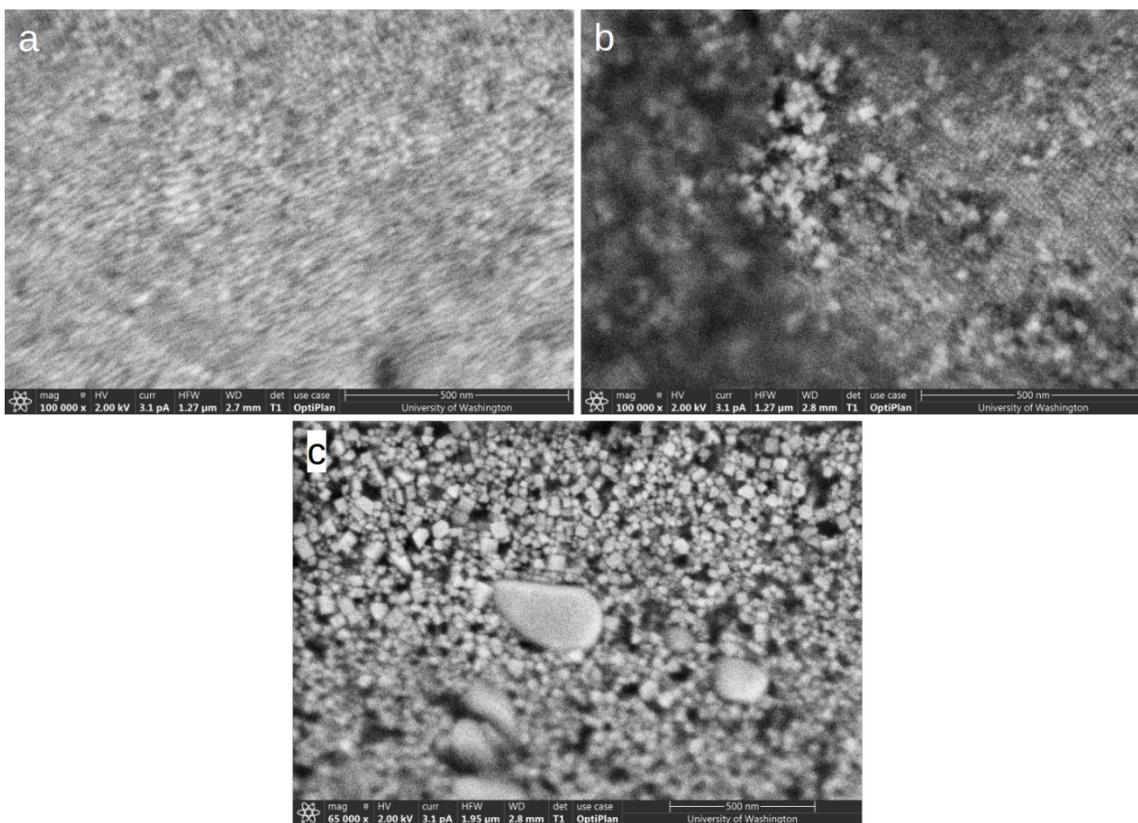

**Figure S19.** SEM images of $Cs_2AgBiI_6$ nanocrystal thin films before (a) and after (b,c) heating (*ex situ*). **(a)** Initial nanocrystal thin film, with nanoparticles generally too small for SEM to resolve. **(b)** Same film after heating to 100°C. Note appearance of some larger particles. **(c)** Film after heating to 125°C. A dominant population of larger particles is now observed.

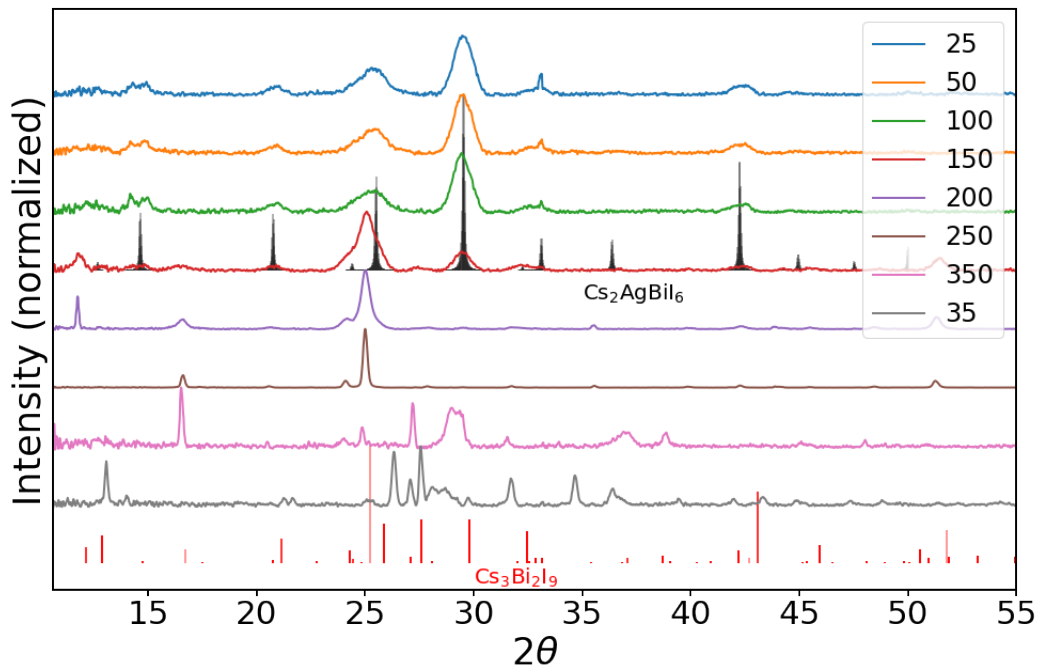

**Figure S20.** *In situ* XRD data for a film of $Cs_2AgBiI_6$ nanocrystals heated in air. The temperature was gradually increased from 25 - 350°C at a rate of 5°C/min using a heating stage, then a pattern was measured at steady temperature for ~10 min. The 35°C data were collected after the previous heating steps. Reference pattern for $Cs_3Bi_2I_9$ taken from ICSD.

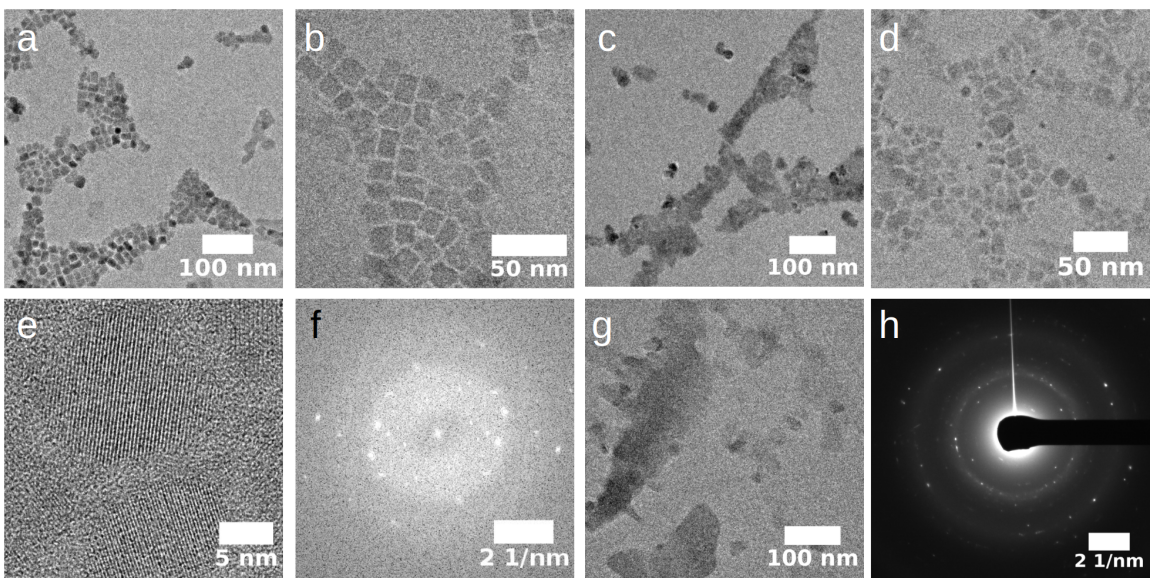

**Figure S21**. TEM images of *in-situ* heating experiments on $Cs_2AgBiI_6$ nanocrystals. Sample was heated to 100°C, held for 10 min, then cooled to 30°C for imaging. Attempts were made to image at high temperature, but rapid beam damage prevented useful data collection. Various products were observed upon heating: **(a)** Varied population of NCs, with some sintering while others appear preserved. **(b)** NCs with preserved morphology after heating. **(c)** NCs showing evidence of sintering/melting after heating. **(d)** NCs with persistent morphology after heating to 120°C. **(e)** HRTEM of preserved $Cs_2AgBiI_6$ NCs. **(f)** FFT of (e), showing elpasolite pattern. **(g)** Area of NCs that fused upon heating. **(h)** Electron diffraction of area shown in (g), showing $Cs_3Bi_2I_9$ hexagonal reflections.

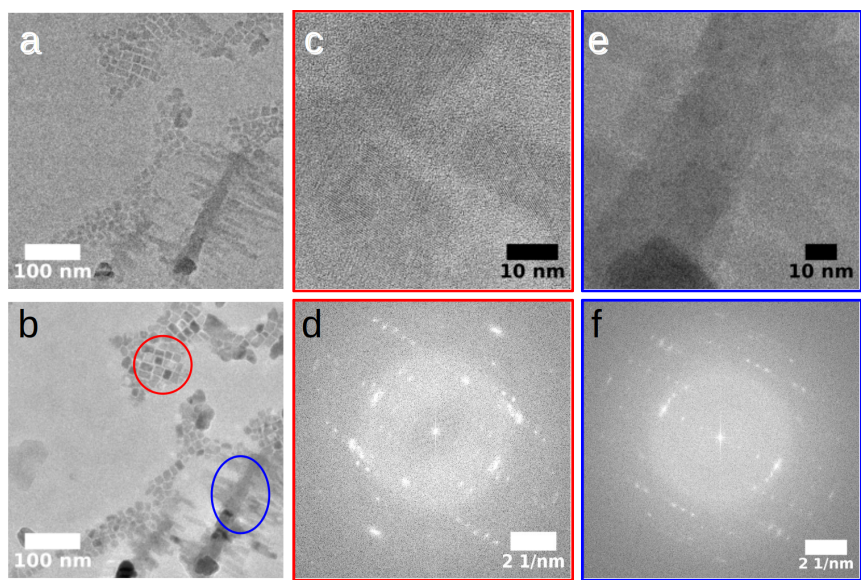

**Figure S22**. Additional TEM images of *in-situ* heating experiments on Cs$_2$AgBiI$_6$ nanocrystals with deliberately broad size distribution, including large branched nanorods. **(a)** Mixed population of Cs$_2$AgBiI$_6$ nanocubes and nanorods prior to heating. **(b)** The same sample area shown in (a) after heating. **(c)** HRTEM of individual NCs post-heating. **(d)** FFT of (c), showing elpasolite pattern. **(e)** HRTEM image of nanorods post-heating. **(f)** FFT of (e), showing elpasolite pattern. We were unable to find evidence of correlation between starting NC size and likelihood of transformation during heating.

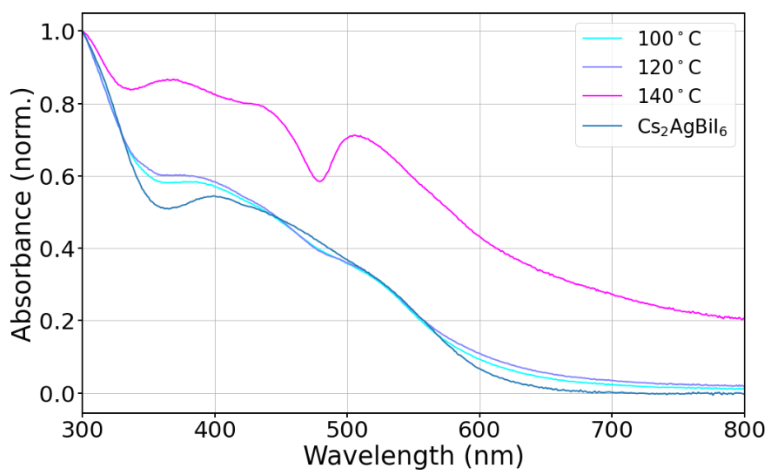

**Figure S23.** Absorption spectra of Cs$_2$AgBiI$_6$ nanocrystals heated in hexanes solution. Heating to 140°C resulted in evaporation of hexanes solvent, even in a sealed container.

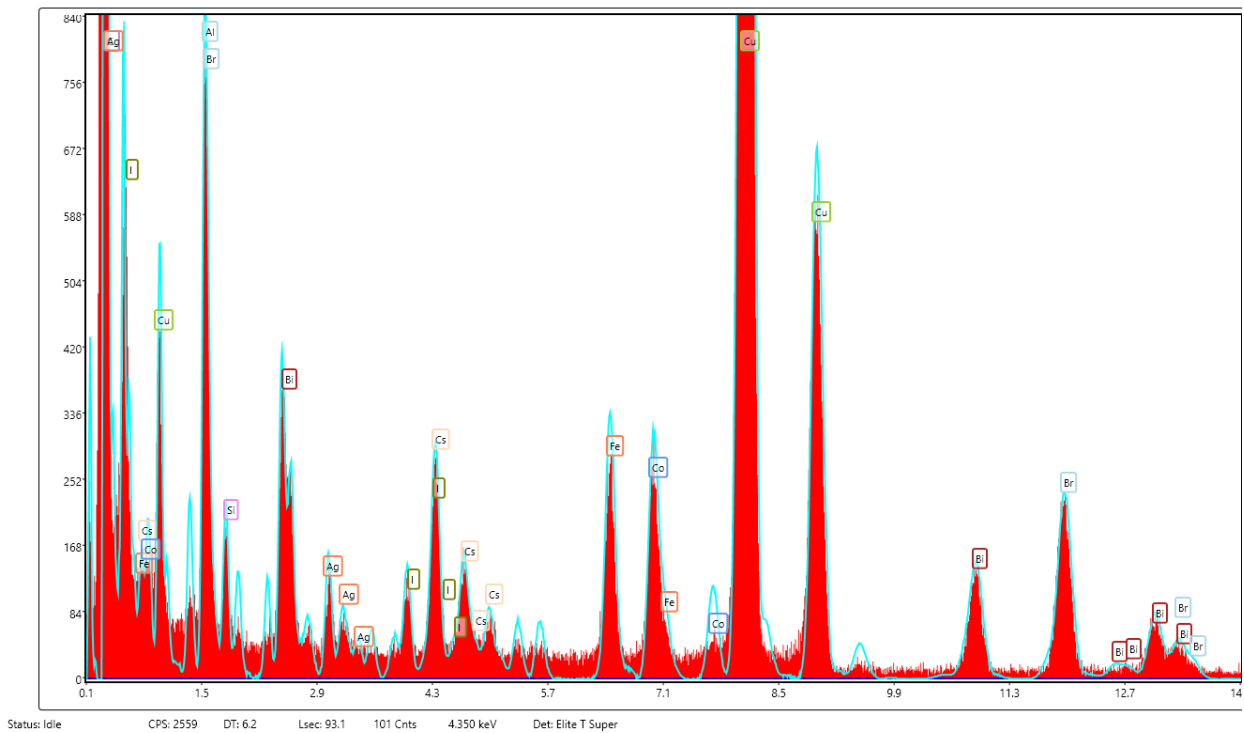

**Figure S24.** TEM EDX data for a representative $Cs_2AgBiX_6$ (X = Br, I) elpasolite NC sample prepared from $Cs_2AgBiBr_6$ NCs by partial anion exchange using TMSI. The data show peaks for Cs, Ag, Bi, Br, and I. The additional peaks are background from the sample grid and holder. EDX (peak intensity ratios) and pXRD both yield the same relative Br:I composition of 1.0:5.0 for this sample.

**References**

(1) Creutz, S. E.; Crites, E. N.; De Siena, M. C.; Gamelin, D. R. Colloidal Nanocrystals of Lead-Free Double-Perovskite (Elpasolite) Semiconductors: Synthesis and Anion Exchange To Access New Materials. *Nano Letters* **2018**, *18* (2), 1118-1123. DOI: 10.1021/acs.nanolett.7b04659.